\documentclass[prb,twocolumn,amsmath,amssymb,superscriptaddress]{revtex4}

\usepackage{graphicx}
\usepackage{dcolumn}
\usepackage{bm}
\usepackage{color}
\usepackage{upgreek}
\newcommand{\ket}[1]{\ensuremath{\left| #1 \right\rangle}}

\begin{document}
\title{Maximising Dynamic Nuclear Polarisation via Selective Hyperfine Tuning }

\author{L. T. Hall}
 \email{liam.hall@unimelb.edu.au}
\affiliation{School of Physics, University of Melbourne, Parkville, VIC 3010, Australia}
\author{D. A. Broadway}
\affiliation{School of Physics, University of Melbourne, Parkville, VIC 3010, Australia}
\affiliation{Centre for Quantum Computation and Communication Technology, School of Physics, University of Melbourne, Parkville, VIC 3010, Australia}
\affiliation{Department of Physics, University of Basel, Klingelbergstrasse 82, Basel CH-4056, Switzerland.}
\author{A. Stacey}
\affiliation{School of Physics, University of Melbourne, Parkville, VIC 3010, Australia}
\affiliation{Centre for Quantum Computation and Communication Technology, School of Physics, University of Melbourne, Parkville, VIC 3010, Australia}
\affiliation{School of Science, RMIT University, Melbourne, Victoria 3001, Australia}
\author{D. A. Simpson}
\affiliation{School of Physics, University of Melbourne, Parkville, VIC 3010, Australia}
\author{J-P. Tetienne}
\affiliation{School of Physics, University of Melbourne, Parkville, VIC 3010, Australia}
\affiliation{Centre for Quantum Computation and Communication Technology, School of Physics, University of Melbourne, Parkville, VIC 3010, Australia}
\author{L. C. L. Hollenberg}
\affiliation{School of Physics, University of Melbourne, Parkville, VIC 3010, Australia}
\affiliation{Centre for Quantum Computation and Communication Technology, School of Physics, University of Melbourne, Parkville, VIC 3010, Australia}

\begin{abstract}
Dynamic nuclear polarisation (DNP) refers to a class of techniques used to increase the signal in nuclear magnetic resonance measurements by transferring spin polarisation from ensembles of highly polarised electrons to target nuclear analytes.
These techniques, however, require the application of strong magnetic fields to maximise electron spin polarisation, limiting pathways for electron-nuclear (hyperfine) spin coupling and transfer.
In this work we show that, for systems of electronic spin $S\geq1$ possessing an intrinsic zero-field splitting, a separate class of stronger hyperfine interactions based on lab-frame cross relaxation may be utilised to improve DNP efficiency and yield, whilst operating at moderate fields. We analytically review existing methods, and determine that this approach increases the rate of polarisation transfer to the nuclear ensemble by up to an order of magnitude over existing techniques. This result is demonstrated experimentally at room temperature using the optically polarisable $S=1$ electron spin system of the nitrogen vacancy (NV) defect in diamond as the source of electron spin polarisation.
Finally we assess the utility of these NV-based approaches for the polarisation of macroscopic quantities of molecular spins external to the diamond for NMR and MRI applications.
\end{abstract}

\maketitle
\section{Introduction}

\begin{figure*}
  \centering
  \includegraphics[width=\textwidth]{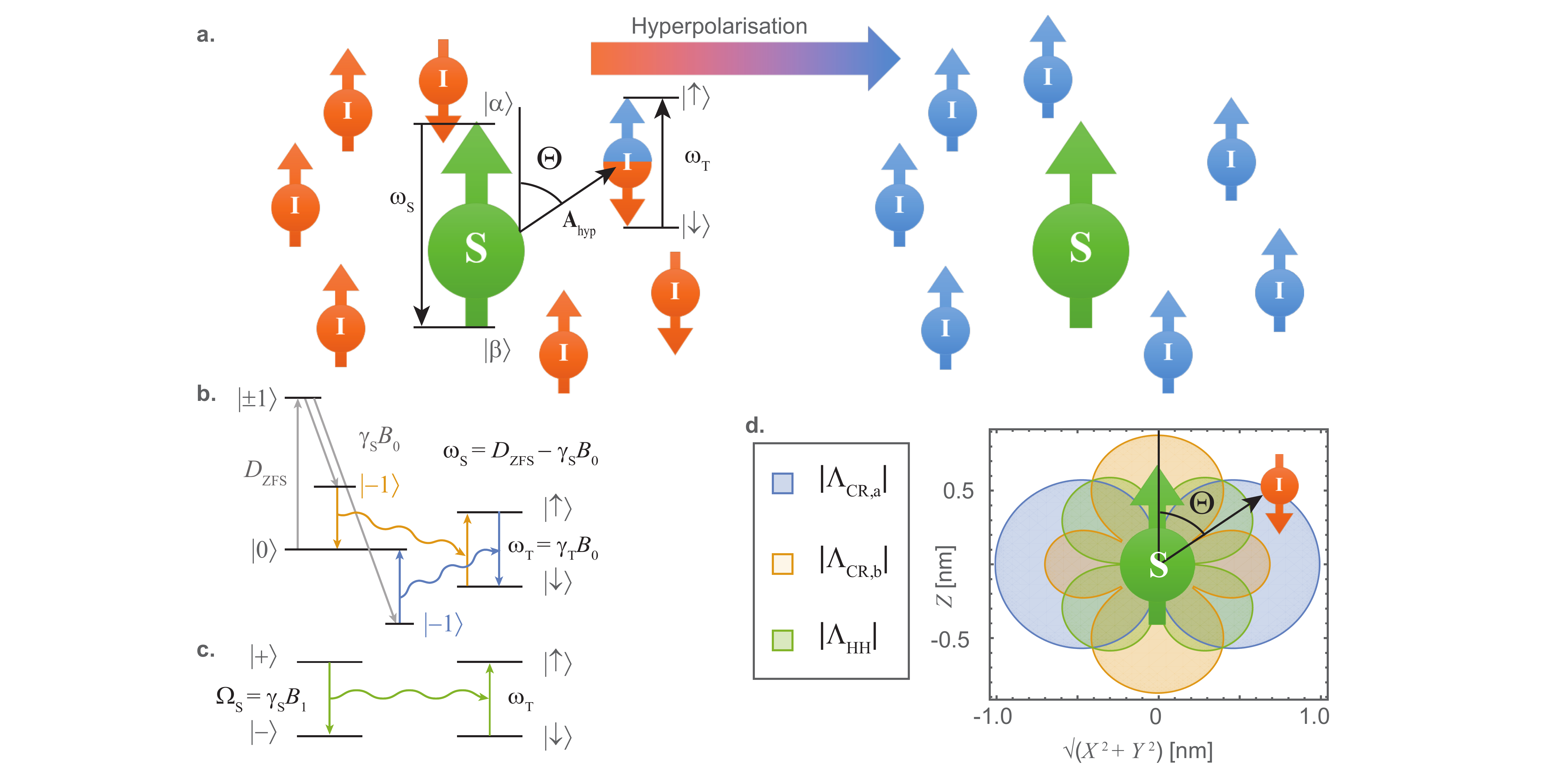}
  \caption{\textbf{Overview of DNP hyperfine mechanisms. a.} Schematic of the electron($\mathbf{S}$)-nuclear($\mathbf{I}$) hyperfine interaction typically employed in solid-effect DNP schemes. Under the appropriate matching conditions, the system precesses between $\ket\alpha\ket\uparrow\leftrightarrow\ket\beta\ket\downarrow$ states. \textbf{b.} The lab-frame cross-relaxation (CR) approach uses the external static field $\left(B_0\right)$ to tune the electron spin's Larmor frequency $\left(\omega_\mathrm{S}\right)$ to that of the nuclear spin $\left(\omega_\mathrm{T}\right)$. Zero field splittings $\left( D_\mathrm{zfs}\right)$ associated with electron spins of $S\geq1$ are utilised to compensate for the mismatch between electron and nuclear gyromagnetic ratios ($\gamma_\mathrm{S}$ and $\gamma_\mathrm{T}$ respectively). Transitions exist before ($\mathrm{CR_b}$) and after ($\mathrm{CR_a}$) the level anti-crossing at $\gamma_\mathrm{S}B_0 = D_\mathrm{zfs}$. \textbf{c.}
  Hartmann-Hahn (HH) based schemes employ an RF field of amplitude $B_1$ to drive the electron spin's Rabi frequency $\left(\Omega_\mathrm{S}\right)$ into resonance with the nuclear Larmor frequency.
  \textbf{d.} Hyperfine lobe structure of the 1\,MHz interaction strength contours of an electron spin (of g-factor $g=2$) coupled to a $^1$H nuclear spin showing the axially symmetric angular dependence of the hyperfine couplings associated with each scheme. }\label{Fig_1}
\end{figure*}

The phenomenon of Nuclear Magnetic Resonance (NMR) has revolutionised our ability to non-invasively investigate the structure of matter on molecular through to anatomical length scales, by exploiting the way in which the magnetic properties of atomic nuclei (nuclear spins) interact with electromagnetic radiation. The sensitivity, and consequently spatial and temporal resolution, of these techniques are dependent on the degree to which the target nuclear spins are mutually aligned, or polarised. In a `brute-force' thermal (or Boltzmann) approach, this polarisation is proportional to the strength of an externally applied magnetic field and inversely proportional to the temperature; however even 400\,MHz ($9\,\mathrm{T}$) spectrometers operating at 1K are only capable of thermal ensemble nuclear polarisations of less than 1\%.

As such, numerous so-called ‘hyperpolarisation’ (HP) approaches have been conceived for the purpose of increasing the fractional excess of nuclear spin polarisation above that at thermal equilibrium. One such approach, referred to as dynamic nuclear polarisation (DNP, Fig.\,1\textbf{a}), takes advantage of the greater degree of spin polarisation present in proximate electron spin baths, $\left\{\mathbf{S}^{(j)}\right\}$, and facilitates the transfer of this polarisation to the hyperfine-coupled nuclear ensemble, $\left\{\mathbf{I}^{(k)}\right\}$, via the application of RF fields. This typically involves the addition of free radicals to increase the size of the electron spin reservoir, as well as the use of strong magnetic fields and cryogenic operation to increase the extent of the initial electron spin polarisation.

One interesting aspect of DNP schemes is the use of RF driving to circumvent the electron-nuclear Zeeman mismatch at moderate-high magnetic fields (Fig.\,1\textbf{b,\,c}). This has the consequence of isolating particular components of their hyperfine interaction, $\mathbf{S}^{(j)}\cdot\mathbf{A}^{(jk)}\cdot\mathbf{I}^{(k)}$, that do not yield the fastest possible transfer of polarisation.
Hartmann-Hahn (HH) techniques, for example, in which the EPR and NMR resonances of the electron and nuclear spins are simultaneously excited with the same Rabi frequency see them interact via the $A_{zz}$ hyperfine component\cite{Sli}. Alternatively, modified Hartmann-Hahn (mHH) techniques\cite{Fri91}, in which the Rabi frequency of the electron is matched to the Larmor frequency of the nuclei, see them instead interact via the $A_{zx}$ and $A_{zy}$ terms.
As the remaining $A_{xx},\,A_{xy},\,A_{xy}$ terms remain unexplored in this context, this prompts the quest for alternative configurations and driving modes, such that the hyperfine interaction (Fig.\,1\textbf{d}) can be re-engineered and maximised for greater nuclear polarisation yields.

In this work, we make use of the remarkable properties of optically polarisable solid state electron spin-triplet systems\cite{Hen90,Tat14} as a convenient source of electron spin polarisation to demonstrate our hyperfine tuning concept. Of these systems, the nitrogen-vacancy (NV) centre in diamond\cite{Doh13} is particularly noteworthy; owing to its convenient room-temperature operation and efficient optical spin polarisation of the electronic ground state, which remove the need for strong magnetic fields and cryogenic operation\cite{Lon13,Alv15,kin15,Sch16,Fer18}. As such, the motivation for employing this system is two-fold: not only does the NV provide an ideal test-bed for this investigation, it may also provide the ideal technology platform for the realisation of compact, cost effective on-chip DNP devices\cite{Abr14,Che16,Fer18,Sch17,Hal17,Tet20}.

We begin with a brief overview of NV-DNP, before narrowing our focus to developing a generalised analytic description of the most promising schemes currently in the literature. We then experimentally compare the performance of these schemes using an ensemble of NV centres in the native $^{13}$C nuclear spin bath in diamond. Finally we use these results to discuss the prospects of scaling-up single-NV
demonstrations of the hyperpolarisation of organic protons external to the diamond under these schemes\cite{Hal17,Sha18} using large ensembles of NVs. Such production of macroscopic ensembles of hyperpolarised nuclear spins external to the diamond would provide cost effective and efficient signal enhancement of analytes in micro-scale NMR or metabolites for functional MRI.

\section{Dynamic Nuclear Polarisation with solid-state electron spins}
The extreme hardware requirements associated with conventional DNP techniques include ultra-strong magnetic and RF fields, cryogenic temperatures and subsequent processing to return to ambient conditions, and the addition of free radicals and catalysts to increase the electron spin reservoir. These requirements add additional costs and complexity that can exceed that of NMR spectrometers or magnetic resonance imaging (MRI) scanners themselves, with lead time and infrastructure costs often precluding all but the most well-funded MRI and NMR facilities from having DNP capability.

Controllable solid-state electron spin systems such as the nitrogen-vacancy (NV) centre defect in diamond, however, offer a possible solution due to their convenient, room temperature source of efficient, optically-induced electron spin polarisation. The NV ground state is an electronic spin triplet that may be efficiently polarised in its $|0\rangle$ state via illumination with green ($\approx$532\,nm) light. These remarkable properties have been harnessed over the last decade to develop a now well-established quantum sensing platform of electromagnetic and thermal phenomena on nanoscopic to macroscopic scales\cite{Ron14,Sch14,Cas18}. Critically, for the purposes of nuclear spin hyperpolarisation, this $>90\%$ efficient\cite{Wal11} NV spin polarisation provides a continuously renewable source of electron polarisation that may be harnessed to polarise proximate ensembles of nuclear spins via NV analogs of existing DNP schemes, including `Overhauser'\cite{Ove53} (OE), `solid'\cite{Jef57} (SE), `cross'\cite{Hwa67} (CE), and `thermal mixing'\cite{Abr78} (TM) effects.

Thermal mixing and cross effect schemes, which utilise both electron-electron and electron-nuclear spin interactions, as part of the DNP process, have been used to transfer polarisation to the native 1.1\% $^{13}$C ensemble also within the diamond crystal. These demonstrations often take advantage of a high-density of both NVs themselves and also nitrogen-donor (P1) centres throughout high-pressure and high-temperature grown bulk diamonds."
Whilst bulk nuclear spin polarisations of up to 6\% (TM)\cite{Kin10} and 1-3\% (CE\cite{Pag18,Hen19}) have been achieved using these methods, as verified by shuttling of the sample into an external NMR spectrometer, polarisation of external spin ensembles are yet to be demonstrated. As the majority of NV centres in such a sample are not close to the diamond surface, polarisation transfer to external targets will necessarily rely on much slower magnetic dipole mediated spin diffusion, rendering it unclear how these methods will translate to efficient polarisation of arbitrary external ensembles of nuclear spins. Methods to directly polarise external nuclei in fluidic and gaseous samples using the Overhauser effect, by taking advantage of stoachastic fluctuations in hyperfine couplings (due to translation and rotational nuclear motion), have been proposed and analysed in detail\cite{Abr14}, but not yet realised.

The DNP methods that have attracted the most attention in the NV community are built upon the `solid effect', in which polarisation of the electron spin is directly and coherently transferred to an adjacent nucleus via their hyperfine coupling (Fig.\,1\textbf{a}).  Whilst protocols harnessing this technique vary considerably, as will be briefly outlined below, the fundamental principle is the same. For an electron spin ($\mathcal{S}$) hyperfine coupled to a single nuclear spin($\mathcal{I}$), the system ($\mathcal{S}\otimes\mathcal{I}$) is initialised in either the \ket{\alpha,\uparrow} or \ket{\alpha,\downarrow} states, where \ket{\alpha} could, for example, refer to one of the electron's Zeeman spin levels $\{\ket{m_s}\}$ (Fig.\,1\textbf{b}), or RF dressed states \ket{+,-} prepared using a resonantly applied RF field (Fig.\,1\textbf{c}); and the $\uparrow,\,\downarrow$ refer to nuclear spin projection parallel and anti-parallel to the external field respectively.
If the system is initially found in the \ket{\alpha,\downarrow} state, it will precess between \ket{\alpha,\downarrow} and \ket{\beta,\uparrow} states; and subsequent re-initialisation of the NV electron will drive the system into the \ket{\alpha,\uparrow} state with some probability. This process is then repeated until the desired degree of nuclear polarisation in the \ket{\uparrow} state is achieved.

In pioneering work by Jacques, et\,al.\cite{Jac09}, this effect was demonstrated under 532\,nm laser illumination by facilitating a cross-relaxation process between the polarised electron spin and its intrinsic $^{15}$N nuclear spin in the electronic excited state, due to the excited state level anticrossing (ESLAC) at $\sim500$\,G. The extent of polarisation was inferred by using optically detected magnetic resonance (ODMR,\cite{Gru97}) to monitor the intensity of its electron paramagnetic resonance (EPR) lines, and was found to be in excess of 99\% for the $^{15}$N nucleus, and in excess of 90\% when similarly applied to a strongly coupled $^{13}$C nucleus in the first coordination shell of the NV. This effect was independently verified in ref.\onlinecite{Fis13} by shuttling a polarised sample into an NMR spectrometer, and showed that an entire $^{13}$C ensemble in a bulk high-pressure, high-temperature (HPHT) dimaond could be polarised to 0.5\%. Recent work suggests that ESLAC interactions with electronic spins in P1 centres may in fact be the dominant mechanism responsible for this technique\cite{Wun17}. Similar methods have been employed by tuning to the ground state level anticrossing between the NV spin and the target ensemble \cite{Wan13}.

In all of the above, optical illumination was applied to the NV in a CW manner, meaning that the electron spin is constantly being driven into its \ket0 state, even during the mixing between \ket{\alpha,\downarrow}\, and \ket{\beta,\uparrow}\, states. Critically, this results in a quantum Zeno-like effect which slows the effective polarisation rate, and consequently limits the spatial extent of the NV-target polarisation effect.
For this reason, pulsed optical methods have been developed to allow for uninterrupted polarisation-transfer interactions limited only by the longitudinal spin relaxation time of the nuclei.

One such approach is to employ precise RF control of the NV spin in order to tune its Rabi frequency to the target nuclear Larmor frequency, in what is known as a modified Hartman-Hahn (HH) or NOVEL pulse sequence (Fig.\,1\textbf{c}). In this approach, the electron is initialised in an eigenstate of the rotating RF field, and its magnetisation is transferred to the lab-frame projection of the nuclear spin via their hyperfine interaction; with the degree and direction of polarisation transfer being dictated via the initial electron state and the interaction time\cite{Lon13}.
Applications of this approach to the native $^{13}$C spin bath within the diamond\cite{Lon13,Sch17} readily saturate the $^{13}$ C nuclei surrounding NV centres\cite{Sch17}, and are currently being explored for the purposes of polarising $^{13}$C spins in diamond nanocrystals\cite{Che15,Joc16,Che16,Ajo18}. This approach is also emerging as a pathway to direct polarisation of nuclei external to the diamond\cite{Fer18,Sha18}.

More recent developments have seen the design of specific pulsed-RF-control polarisation sequences that improve the electron spin's robustness to environmental noise, field alignment, and pulse errors; at the cost of slight reductions in the effective hyperfine coupling as compared with the ideal HH scenario. These include approaches to introduce controlled over-rotations (PolCPMG\cite{Lan19}), time delays (TOP-DNP\cite{Tan20}), and precise engineering of the electron spin Hamiltonian (PulsePol\,\&\,PolXY\cite{Sch18}) to dynamical decoupling protocols in such a way that the hyperfine dynamics mimic the HH protocol. However under many scenarios of practical interest these techniques do not address, and in some cases amplify, the polarisation-limiting processes present in these systems. The fundamental limits to these engineered schemes and comparison with alternative approaches will be a major focus of this work.

Alternatively, the use of RF fields may be avoided by using an external magnetic field to tune the transition frequencies of the NV to that of the target (Fig.\,\ref{Fig_1}\,\textbf{b}, orange, blue), thereby facilitating a direct lab-frame cross-relaxation (CR) exchange of polarisation.
This approach is achievable for electronic spin systems which exhibit decreasing transition energies under the Zeeman effect; a feature of all spin $S>1$ systems which having appreciable zero-field spin-spin interactions
 As noted in ref.\,\onlinecite{Jac09} however, this method requires precise control over both the strength and alignment of this field. Owing to a precise understanding of the NV GSLAC structure\cite{Bro16}, which occurs in the vicinity of the fields required for DNP, CR based NV-DNP of both proximate $^{13}$C and external $^{1}$H have been realised\cite{Hal17}, a feat that has since been also realised using the NOVEL sequence\cite{Sha18}.

In this work, we study the fundamental differences between Hartmann-Hahn and cross-relaxation approaches (including pulsed variations thereof) in order to understand their underlying limitations and potential for nuclear spin hyperpolarisation.
In Section\,\ref{BulkC13Sec}, we begin with an analytic treatment of the these techniques, and show that appreciable differences in their predicted nuclear polarisation rates arise from their hyperfine coupling geometries (\ref{Sec_bulk_HF}). Using the NV system as an example, we apply this framework to study the DNP case in which the source of electronic polarisation is immersed in the bulk of the nuclear spin ensemble, as represented by the $^{13}$C nuclear spin bath within the diamond crystal (\ref{Sec_bulk_stat}); and experimentally demonstrate for a range of NV centres that the polarisation buildup under the CR approach exceeds that of HH by an order of magnitude (\ref{Subsec_13C_Results}). Finally, in Section\,\ref{External_prospects}, we apply our framework to the case in which the electron and nuclear spin ensembles are spatially separated, as represented by the $^1$H proton spins of organic polymers residing on the diamond surface, and reexamine the prospects for NV-DNP of external nuclear spin ensembles. We show that, despite the engineering of RF based schemes to decouple the NV spin from sources of dephasing, the dephasing and motion of target nuclear spins lead to reductions in polarisation rates that significantly undermine the advantages in electron spin coherence offered by these schemes.

\section{Hyperpolarisation from within a bulk nuclear spin ensemble}\label{BulkC13Sec}
In addition to the interest surrounding the direct polarisation of nuclear spins external to the diamond surface, spin polarisation of the bulk nuclei within the diamond has many important applications. These include the production of spin polarised $^{13}$C within diamond nanocrystals; and constitute the first step of prospective methods to transfer bulk nuclear polarisation via nuclear-nuclear coupling with external spins. In addition, this system provides an important experimental model system to study the differences between the DNP schemes under the scenario of free-radical electron spins embedded within the bulk of a nuclear spin bath: a scenario typically encountered in conventional DNP associated many NMR experiments.

As such, in this section we first discuss the statistical distributions of hyperfine couplings and polarisation yields in bulk systems for the protocols introduced above. We subsequently make a direct experimental comparison of CR and HH techniques in the context of their relative abilities to polarise the $^{13}$C spins that comprise 1.1\% of carbon atoms within the diamond crystal, and confirm that the total hyperfine coupling of each NV is indeed an accurate quantitative predictor of its ability to polarise ensembles of surrounding nuclei.

We first briefly describe the fundamental theoretical framework of the techniques noted above in the context of a single target spin, then move on to consideration of the polarisation of multi-spin target ensembles.
Using the NV centre in diamond as an example, the CR and HH (and pulsed variations thereof) polarisation schemes are concerned with the transfer of the optically induced electron spin polarisation of the NV centre to adjacent nuclear spins.  In both cases, we solve for the evolution of the combined NV\,+\,target density matrix according to the von Neumann equation,
\begin{eqnarray}
  \frac{\mathrm{d}}{\mathrm{d}t}\rho(t) &=& -i\left[\mathcal{H}(t),\rho(t)\right] + \mathcal{L}(\rho,t),\label{Eq_Lindblad}
\end{eqnarray}
 where
 \begin{eqnarray}
  \mathcal{H} &=& \mathcal{H}_S+\mathcal{H}_I+\mathcal{H}_\mathrm{int},
\end{eqnarray}
describes the effect of the Hamiltonians of the spin, target, and their interaction respectively. The Lindbladian term, $\mathcal{L}$, describes competing incoherent effects such as decoherence and relaxation, however their detailed treatment will be left until the consideration of external nuclei in Section\,\ref{External_prospects}, as they are not appreciable in the present context.

We refer to nuclear spin polarisation in regards to its magnetic sub-levels as defined by interaction with an external magnetic field,
\begin{eqnarray}
  \mathcal{H}_I &=& \sum_j\omega^{(j)}_\mathrm{T} \mathcal{I}^{(j)}_z.
\end{eqnarray}
 For simplicity we restrict our discussion to $I=\frac{1}{2}$ systems, although this framework may be generalised to include additional effects such as quadrupole interactions for cases of $I\geq1$. The electron spin Hamiltonian, $\mathcal{H}_S$, will depend on the RF pulse sequence applied to the electron spin in the polarisation scheme under consideration.

\subsection{Hyperfine dynamics}\label{Sec_bulk_HF}
\subsubsection{Single-spin target environments}
The interaction between the NV and a given target spin are dictated by the magnetic dipole hyperfine interaction, as described by
\begin{eqnarray}
  \mathcal{H}_\mathrm{int} &=& \mathbf{S}\cdot\mathbf{A}\cdot\mathbf{I},
\end{eqnarray} where $\mathbf{S}=\left(\mathcal{S}_x,\mathcal{S}_y,\mathcal{S}_z\right)$, $\mathbf{I}=\left(\mathcal{I}_x,\mathcal{I}_y,\mathcal{I}_z\right)$ are the vector spin operators for the electron and target nucleus respectively. The hyperfine tensor, $\mathbf{A}$, dictates their mutual interaction according to
\begin{eqnarray}
  \mathbf{A} &=& \left(
                    \begin{array}{ccc}
                      A_{xx} & A_{yx} & A_{zx} \\
                      A_{yx} & A_{yy} & A_{zy} \\
                      A_{zx} & A_{zy} & A_{zz} \\
                    \end{array}
                  \right),\\
                  A_{mn} &=& \frac{a}{{R}^3}\left(\delta_{mn} - 3\frac{R_m R_n}{r^2}\right),
\end{eqnarray} where $\mathbf{R}=(R_x,R_y,R_z)$ is the spatial separation vector, and $a=\frac{\mu_0\hbar}{4\pi}\gamma_\mathrm{NV}\gamma_\mathrm{T}$ defines the magnitude of the coupling strength in terms of the gyromagnetic ratios of the NV and target spins, $\gamma_\mathrm{NV}$ and $\gamma_\mathrm{T}$, respectively.

In the HH/Novel sequence (Fig.\,1\textbf{c}), the NV spin is optically initialised in its $|0\rangle$ state, and then prepared in one of the superposition states $|\pm\rangle\equiv\frac{1}{\sqrt2}\left(|0\rangle \pm|-1\rangle\right)$ using a $\pi/2$ microwave (MW) pulse tuned to its $|0\rangle\leftrightarrow|-1\rangle$ transition. A second microwave field, also tuned to this transition but phase shifted by $\pi/2$, is introduced; however its microwave power is chosen such that the Rabi frequency of the NV spin ($\Omega_\mathrm{S}$) matches the Larmor frequency of the target spin ($\omega_\mathrm{T}$). In the absence of any other disturbance, this `locks' the NV spin vector into alignment with the rotating MW field, thereby defining its quantisation axis. When the HH resonance condition ($\Omega_\mathrm{S}=\omega_\mathrm{T}$) is achieved, the $A_{zx}$ and $A_{zy}$ components of their hyperfine interaction will drive the NV spin between its $|+\rangle$ and $|-\rangle$ states, and the target between its $|\uparrow\rangle$ and $|\downarrow\rangle$ states.

The control-pulse based sequences noted above (ie PolCPMG, TOP-DNP, polXY, PulsePol) are designed to mimic this behaviour, but do so by instead periodically flipping the NV spin about the rotating $x$ and/or $y$ axes associated with the microwave field. As such, these schemes employ the same hyperfine components as the HH case, but with a slight reduction in coupling strength\cite{Sch18}. As such, we thus regard control-pulse based protocols as part of the HH suite, and specific details in which their respective robustnesses may lead to further differences in polarisation rates will be given where necessary.

In the case of CR, a static magnetic field is used to bring the transition frequencies of the NV and target spins into resonance, allowing for a direct exchange of polarisation via the $A_{xx},\,A_{xy},\,\mathrm{and}\,A_{yy}$ components of their hyperfine interaction\cite{Hal16}. Owing to the zero field splitting of the NV spin, two such resonances exist in principle, either side of the NV ground-state level anti-crossing (GSLAC)(Orange and blue, Fig.\,1\textbf{b}).
In practice however, the GSLAC details create some resonance 'dead zones'. For example,
in the case of a $^{14}$NV coupled to proton spins, only the after-GSLAC interaction is observed due to the avoided crossing before the GSLAC, however both transitions are observed for nuclei with smaller gyromagnetic ratios. In the case of $^{15}$NV, no transition frequency exists below roughly 3\,MHz, making it suitable for NV-DNP of  $^{1,3}$H and  $^{19}$F spins, but little else (detailed discussions of these considerations may be found in refs.\,\onlinecite{Bro16,Woo17}). As proton spins are a highly desirable external target for NV-DNP, we consider both before and after GSLAC cases (denoted CR$_\mathrm{b}$ and CR$_\mathrm{a}$ respectively) in the following discussion. However we note that a practical device for the polarisation of arbitrary species would employ the CR$_\mathrm{a}$ transition of a $^{14}$NV system. Moreover, as will be discussed below, the CR$_\mathrm{a}$ case utilises the strongest NV-nuclear coupling terms, thus presenting scope for the greatest rate of polarisation transfer.

Regardless of the polarisation scheme employed, a single environmental target (T) spin-$\tfrac{1}{2}$ system  will be driven between its $\left|\uparrow\right\rangle$ and $\left|\downarrow\right\rangle$ states by virtue of its coupling to the NV electron spin. Its respective polarisation after some interaction time $t$ is described by
\begin{eqnarray}
  P_\mathrm{T}(t)&=& P_\mathrm{T}(0) - \left(P_\mathrm{T}(0)+\frac{1}{2}\right) 
    \sin^2\left(\frac{\Lambda t}{2}\right)\label{PolNoDephasing},
\end{eqnarray}where $P_\mathrm{T}(0) $ is its initial polarisation at $t=0$, and $\Lambda$ is the effective polarisation rate due to the scheme-specific hyperfine coupling between it and the NV spin.
We note that for the control-pulse-based schemes, there are some subtleties associated with how the interaction time should be chosen in order to relate to the nuclear Larmor frequency,
however under the appropriate choice, the nuclear polarisation will follow the same dependence as given in Eq.\,\ref{PolNoDephasing}. We note also that this expression does not hold for cases in which the NV $\left(\Gamma_\mathrm{NV}\right)$ or target $\left(\Gamma_\mathrm{T}\right)$ spin dephasing exceeds the hyperfine coupling $\left(\Lambda\right)$; such cases are treated in the following section.

Where the various polarisation schemes differ most prominently is in the angular dependencies of their respective polarisation rates, which lead to remarkably different polarisation rates and field geometries for targets both within and external to the diamond crystal,
\begin{eqnarray}
 \Lambda_\mathrm{CR,a}&=& \frac{1}{\sqrt{2}}\sqrt{\left(A_{xx}- A_{yy}\right)^2+\left(A_{xy}+A_{yx}\right)^2},\nonumber\\
 &=&\frac{3 a}{\sqrt{2} R^3} \sin ^2(\Theta ),\nonumber\\
 \Lambda_\mathrm{CR,b}&=& \frac{1}{\sqrt{2}}\sqrt{\left(A_{xx}+ A_{yy}\right)^2},\nonumber\\
 &=&\frac{ a}{\sqrt{2} R^3}\left| 3\cos ^2(\Theta )-1\right|,\nonumber\\
   \Lambda_\mathrm{HH}&=& \frac{1}{2}\sqrt{A_{zx}^2+ A_{zy}^2},\nonumber\\
  &=&\frac{3 a }{2 R^3}\left|\sin (\Theta ) \cos (\Theta )\right|,\label{LambdaRTheta}
\end{eqnarray}where $R$ is the distance between the NV and target, and $\Theta$ is the polar angle of this separation. The spatial dependencies of $\Lambda$ for these 3 cases are shown in Fig.\,\ref{Fig_1}\,\textbf{d}.
In addition, the effective hyperfine coupling under pulse-based variations of HH will necessarily be weaker than the passive case given above, with the PolCPMG (PC) and PulsePol (PP) cases
given respectively by
\begin{eqnarray}
 \Lambda_\mathrm{PC}&=& \frac{2}{\pi}\Lambda_\mathrm{HH},\nonumber\\
   \Lambda_\mathrm{PP}&=& \frac{2  }{3\pi}(2+\sqrt{2})\Lambda_\mathrm{HH}.\label{LambdaPulses}
\end{eqnarray}
Having established the relative hyperfine coupling strengths, the subsequent analysis of the environmental polarisation processes associated with each technique proceeds identically, as it depends only on the choice of $\Lambda$. As such, we will omit the subscripts in the analysis that follows.

By utilising the NV spin state as a proxy for NV-nuclei interactions, we may optimise the chance of a flip-flop occurring.
As will be demonstrated experimentally in Section \ref{Subsec_13C_Results}, choosing an interaction time of $t=\pi/\Lambda\equiv\tau$, ensures maximal transfer of the optically induced NV polarisation to the intended target.

\begin{figure*}
  \centering
  \includegraphics[width=\textwidth]{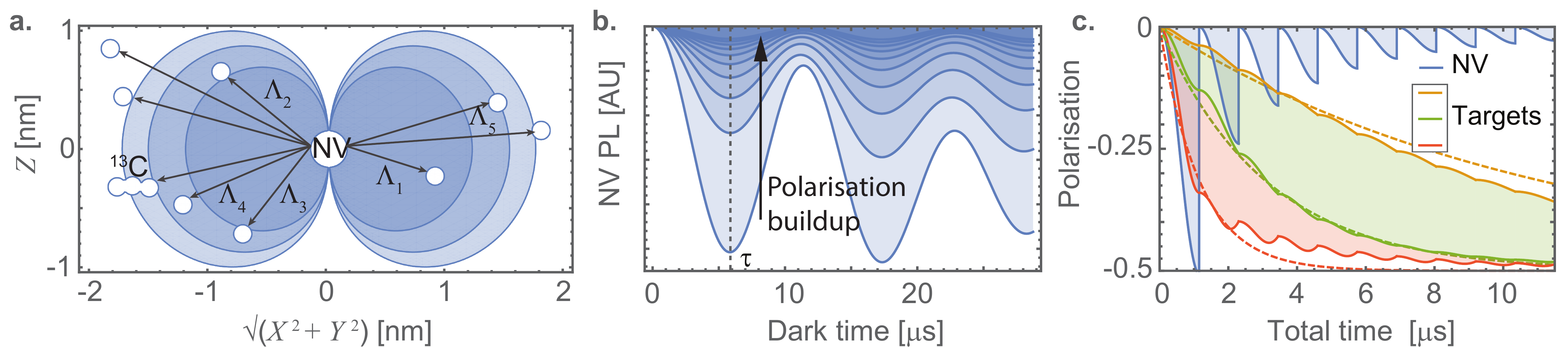}
  \caption{\textbf{Theoretical overview of NV-DNP mechanism for the case in which target nuclei may be treated as discreet entities; using the native 1.1\% $^{13}\mathrm{C}$ bath in diamond under lab-frame cross-relaxation as an example.} \textbf{a.} Spatial distribution of a particular realisation of possible $^{13}$C locations and couplings $\left(\left\{\Lambda_j\right\}\right)$ surrounding the NV. Contours depict couplings of $\Lambda = $ 100\,kHz, 50\,kHz, and 33\,kHz, corresponding to interaction times of $\tau = $ 10\,$\upmu$s, 20\,$\upmu$s, and 30\,$\upmu$s respectively.
  \textbf{b.} Simulated NV PL arising from coherent interactions with the nuclei shown in \textbf{a}, for increasing levels of initial nuclear polarisation from an initial mixed environmental state. A local minimum in the time-dependent fluorescence persists as the environment is polarised, and is used to define the interaction time, $\tau$. As the polarisation of the surrounding nuclei increases, the visibility of coherent dynamics in the NV PL decreases.
  \textbf{c.} Theoretical model of NV/target polarisation dynamics. The inability of the NV to tunnel out of its initial state (blue) is characteristic of target polarisation.
  Dashed lines denote the analytic expression given in Eq.\,\ref{Exp_decay} for pure hyperfine cooling. Deviations of the exact numerical result from pure exponential behaviour are due to hyperfine mediated spin diffusion. This effect is negligible on the timescales studied in this work.}\label{Fig_2}
\end{figure*}

\subsubsection{Many-spin target environments}
\begin{figure}[b]
  \centering
  \includegraphics[width=\columnwidth]{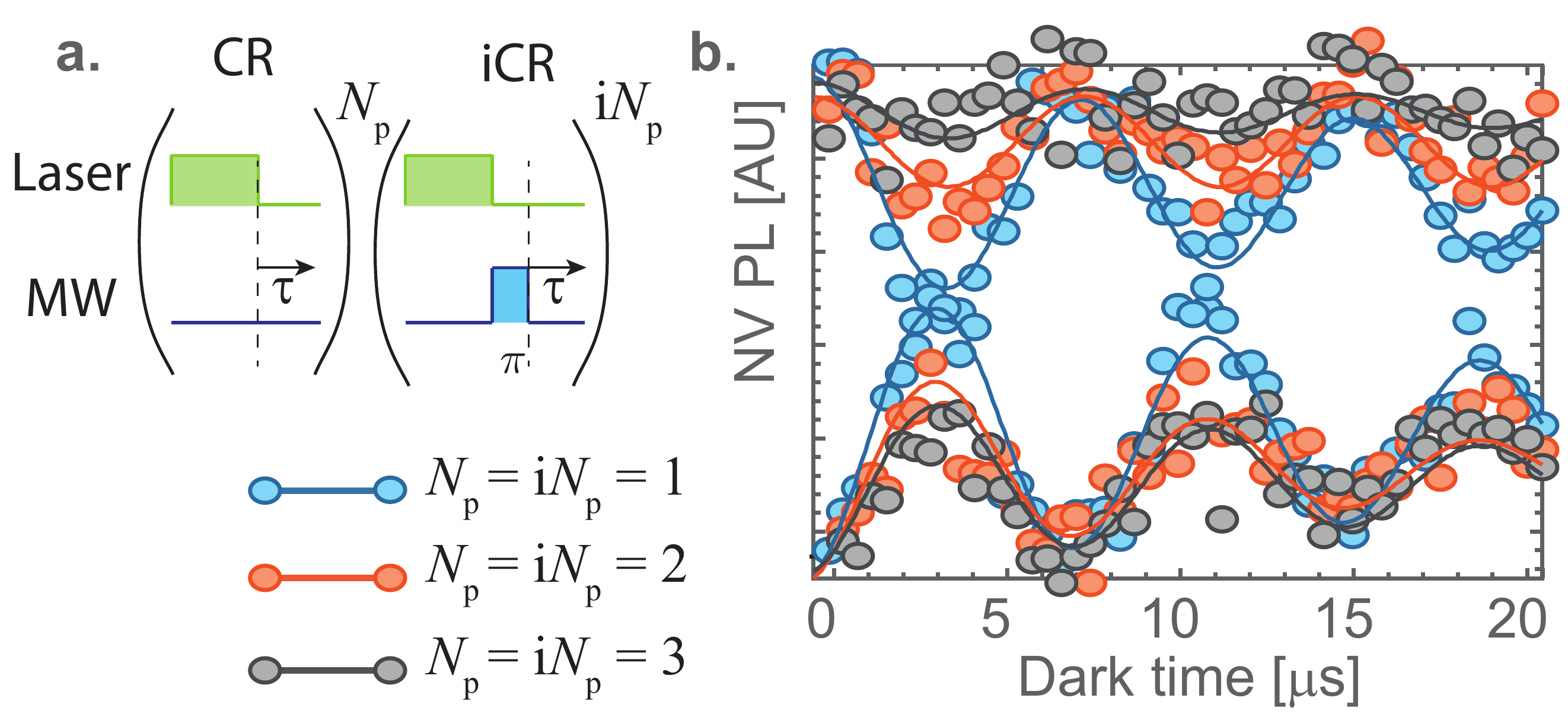}
  \caption{\textbf{Experimental verification of pulse scheme.}
  \textbf{a.} Pulse sequence used to control polarisation buildup. Depolarisation is achieved via an alternating CR-iCR $\left(N_\mathrm{p}=\mathrm{i}N_\mathrm{p}=1\right)$ sequence, whereas  buildup occurs for $N_\mathrm{p}=\mathrm{i}N_\mathrm{p}\geq2$.
   \textbf{b.} Experimental verification of the scheme in \textbf{a} using $N_\mathrm{p}=\mathrm{i}N_\mathrm{p}=\left\{1,2,3\right\}$ (blue, black, and red data, respectively; solid lines represent fits to an exponentially decaying cosine).}\label{Fig_3a}
\end{figure}

For environments containing multiple target spins, the optimal interaction time is still $\tau = \pi/\Lambda$ as above, following a substitution of
\begin{eqnarray}
  \Lambda &\mapsto& \sqrt{\sum_j\Lambda_j^2},\label{LambdaDef}
\end{eqnarray}
where the $\Lambda_j$ refers to the hyperfine coupling between the NV and $j^\mathrm{th}$ bath spin (Fig.\,\ref{Fig_2}\textbf{a}). Subsequent polarisation of the target ensemble will reduce the NV's ability to tunnel out of its initial state; however $\tau$ persists as a local minimum in the NV signal, and hence the optimal choice of interaction time. 
This is illustrated by plotting the population of the NV's initial state as a function of $\tau$ for the case of an NV centre in its native 1.1\% $^{13}$C bath in Fig.\,\ref{Fig_2}\,\textbf{b} (see figure caption for details).
Experimental verification of this approach is detailed later in this section (Sec.\,\ref{Subsec_13C_Results}; and Fig.\,\ref{Fig_3a}).

\begin{figure*}
  \centering
  \includegraphics[width=\textwidth]{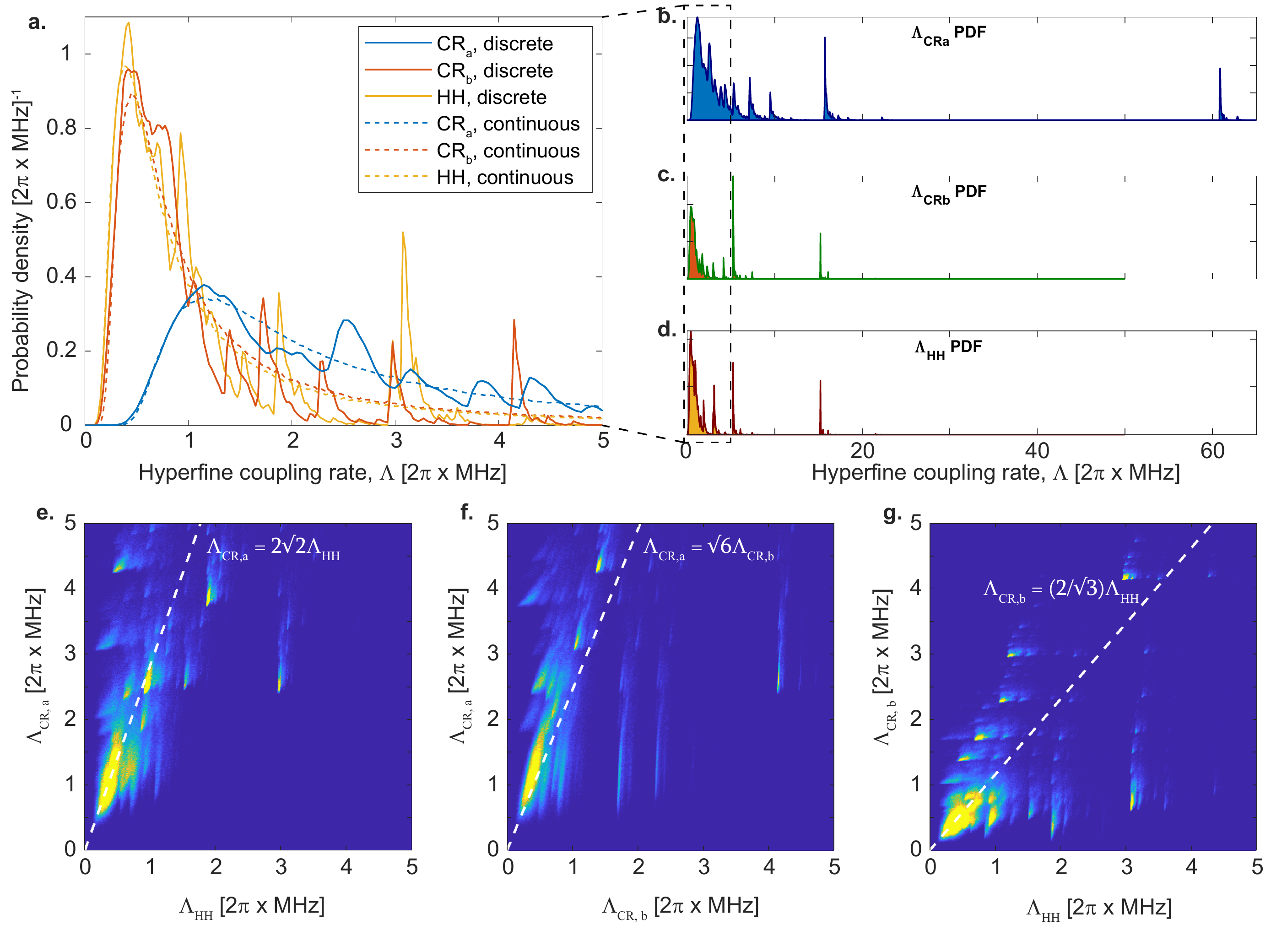}
  \caption{\textbf{Study of hyperfine coupling distributions. a. }Zoomed probability density functions for the realisation of total hyperfine couplings due to randomly distributed $^{13}$C nuclei around a central NV centre. Solid curves arise from nuclei only being allowed to adopt finite lattice positions in the diamond crystal. Dotted curves arise from allowing nuclei to adopt any position. \textbf{b-d.} Probability distributions from \textbf{a} illustrating their full support. \textbf{e-g.} Joint probability distributions showing the relationship between each scheme's hyperfine coupling realisations. White dashed lines depict the solid-angle averaged ratios between the total hyperfine couplings given in Eq.\,\ref{LambdaRTheta}.
  Projections of these distributions along each respective axis gives the PDFs depicted in \textbf{a-d}. Statistical parameters: averages taken over 10$^6$ random realisations, with a radial cutoff of 4\,nm.}\label{Fig_prob_dist}
\end{figure*}

Under this choice of $\tau$, the resulting change in polarisation of spin $j$ due to the hyperfine cooling is given by
\begin{eqnarray}
  P_j(t+\tau) - P_j(t)&=&  -\left[P_j(t)+\frac{1}{2}\right]\frac{\Lambda_j^2}{\Lambda^2}\nonumber\\
  &&-\sum_k\left[P_j(t)-P_k(t)\right]\frac{\Lambda_j^2\Lambda_k^2}{\Lambda^4},\,\,\,\,\,\,\label{dynamical_system}
\end{eqnarray}where the final term arises due to NV-mediated interactions between target spins (see, for example, ref.\,\onlinecite{Cyw09} for a detailed general discussion of these effects, and ref.\,\onlinecite{Hal17} for their origin in the present context). By repeated application of these NV-DNP sequences, we may build up the polarisation of the environment via repeated transfer of the NV spin polarisation.

Additional processes due to dipole coupling between environmental spins, and their own spin-lattice or longitudinal relaxation need also be considered; and, as each successive step in this series takes place over a timescale of $\tau$, this constitutes an effective dynamical system describing the polarisation of the nuclear spins $\{P_j\}$,
\begin{eqnarray}
  \frac{\mathrm{d}}{\mathrm{d}t}P_j(t) &=&-u_j\left[P_j(t)+\frac{1}{2}\right]-\Gamma_1P_j(t)\nonumber\\
  && -\sum_k\left(D_A^{jk}+D_B^{jk}\right) \left[P_j(t)-P_k(t)\right],\,\,\,\label{DEDisc}
\end{eqnarray}where $D_{A,B}^{jk}$ are diffusion terms due to hyperfine-mediated and dipole-mediated interactions, respectively, $\Gamma_1$ is the longitudinal spin relaxation rate of the target spins, and
\begin{eqnarray}
  u_j &=& \frac{\Lambda_j^2}{\pi\Lambda},
  \label{Eq_cool_rate_coherent}
\end{eqnarray}is the effective hyperfine cooling rate. The results of a simulated example realisation of this system are presented in Fig.\,\ref{Fig_2}\textbf{c}.

The dynamics experimentally observed in the study of $^{13}$C nuclei coupled to NV centres deep within the diamond bulk are dominated by the hyperfine cooling terms, $u_j$. In the absence of all other processes, solution of Eq.\,\ref{DEDisc} shows that target polarisations exhibit a purely exponential saturation, 
\begin{eqnarray}
  P_j(t) &=&  P_j(0) e^{-t \left(\Gamma _{1}+u_j\right)} \nonumber\\
  &&- \frac{1}{2}\frac{u_j}{ \Gamma_1 +u_j}\left[1-e^{-t \left(\Gamma _{1}+u_j\right)}\right]\label{Exp_decay}
\end{eqnarray} thereby achieving a steady state polarisation given by
\begin{eqnarray}
  P_j(t)\biggr|_{t\gg{1}/{\Gamma _{1}}} &\sim&  - \frac{1}{2}\frac{u_j}{ \Gamma_1 +u_j},\label{no_deph_SS}
\end{eqnarray} where the total polarisation imparted is then the sum over that of all individual nuclei. This expression details the occurrence of local polarisation saturation, which leads to inefficient macroscopic polarisation unless some form of polarisation transport is introduced. This could be in the form of spatial or magnetic dipole-mediated diffusion, or convection; which in-turn necessitates an understanding of the effects of both target motion and non-target paramagnetic impurities on the polarisation process described above. These considerations will thus be deferred until Section\,\ref{External_prospects}.

\subsection{Analysis of hyperfine statistics}\label{Sec_bulk_stat}
As discussed above, the total hyperfine coupling felt by the central NV spin is the primary measurable quantity determining the rate of polarisation transfer to the target ensemble. Therefore, prior to discussing the experimental performance comparison of the CR and HH schemes, it is instructive to consider the distribution of possible total hyperfine field strengths felt by a central NV spin due to a random spatial distribution of surrounding $^{13}$C nuclei. This not only allows us to investigate the average polarisation rates expected from an ensemble of NV centres, but also affords a statistical basis for interpreting the multiple cases of single-NV experiments considered below.

In order to compare the statistical distributions of the expected hyperfine field strengths associated with each scheme, $\left(\mathrm{P}_\mathrm{CR,a}\left(\Lambda_\mathrm{CR,a}\right)\right.$, $\mathrm{P}_\mathrm{CR,b}\left(\Lambda_\mathrm{CR,b}\right)$, and $\left.\mathrm{P}_\mathrm{HH}\left(\Lambda_\mathrm{HH}\right)\right)$ in a diamond of natural $^{13}$C abundance, we simulated $10^6$ realisations of randomly distributed $^{13}$C nuclei occupying diamond lattice positions around a central NV centre, and evaluated the total hyperfine coupling according to Eq.\,\ref{LambdaDef}. The NV was aligned along the $\langle111\rangle$ crystallographic axis (taken to be the $z$-axis), with sites at $(0,0,0)$ and $(0,0,1.54{\mathrm{\AA}})$ allocated to the nitrogen and vacancy of the NV respectively. All other sites had a 1.1\% chance of being occupied by a $^{13}$C nucleus. The radial cutoff was set to 4.2\,nm (12 unit cells), allowing for a total of 54,253 lattice sites, and an average  population of $\sim597\,^{13}$C nuclei for each NV; thereby affording a spectral resolution of $\sim$300\,Hz in the binning of hyperfine couplings.

The resulting distributions (Fig.\,\ref{Fig_prob_dist}.\,\textbf{a},\textbf{b}-\textbf{d}) show that realisations of CR$_\mathrm{a}$ couplings are more heavily weighted to higher values of $\Lambda$ than both CR$_\mathrm{b}$ and HH cases. At the very large-$\Lambda$ extremity of each distribution, the support becomes very sparse, reflecting the close proximity to the NV and limited availability of lattice sites.

In order to understand the effect of $^{13}$C spins only being able to occupy discrete positions within the lattice, we also studied the distribution of hyperfine couplings for the same average density, but under the scenario of the nuclei being able to adopt any position relative to the NV (dashed curves, Fig.\ref{Fig_prob_dist}.\,\textbf{a}.). The comparison shows the same leading order (small-$\Lambda$) behaviour as the discreet-site case, however a long continuous tail replaces the peaks associated with nearby discrete lattice positions. We note that the continuum distribution is reflective of conventional DNP experiments in which ensembles of DNP radicals and target nuclei may reside in any relative position.

The scaling of both the leading order and tail sections of these distributions may be understood by examining the hyperfine coupling distribution from a single $^{13}$C, as previous work\cite{Hal16} has shown that, on average, approximately 70\% of the central spin relaxation due to a bulk distribution of surrounding spins may be attributed to the nearest spin. For an average bulk density $n$, the probability distribution associated with the distance to the closest spin is given by
\begin{eqnarray}
  \mathrm{P}_R(R) &=& {4\pi n R^2}\exp\left(-\frac{4\pi}{3}nR^3\right).
\end{eqnarray} A change of variables to $\Lambda^{(\mathrm{s})}\equiv a^{(\mathrm{s})} /R^3$, where $a^{(\mathrm{s})}$ sets the effective angular coupling for the scheme in question, yields the probability distribution for $\Lambda$,
\begin{eqnarray}
  \mathrm{P}_\Lambda(\Lambda) &=& \frac{4 \pi  a n }{3 \Lambda^2} \exp\left({-\frac{4 \pi  a n}{3 \Lambda}}\right),
\end{eqnarray}which exhibits a $1/\Lambda^2$ tail, and a very abrupt cutoff for small $\Lambda$, reflecting the fact that the chance of not finding a $^{13}$C spin in a volume surrounding an NV is exponentially suppressed as the size of that region is increased.

\begin{figure*}
  \centering
  \includegraphics[width=\textwidth]{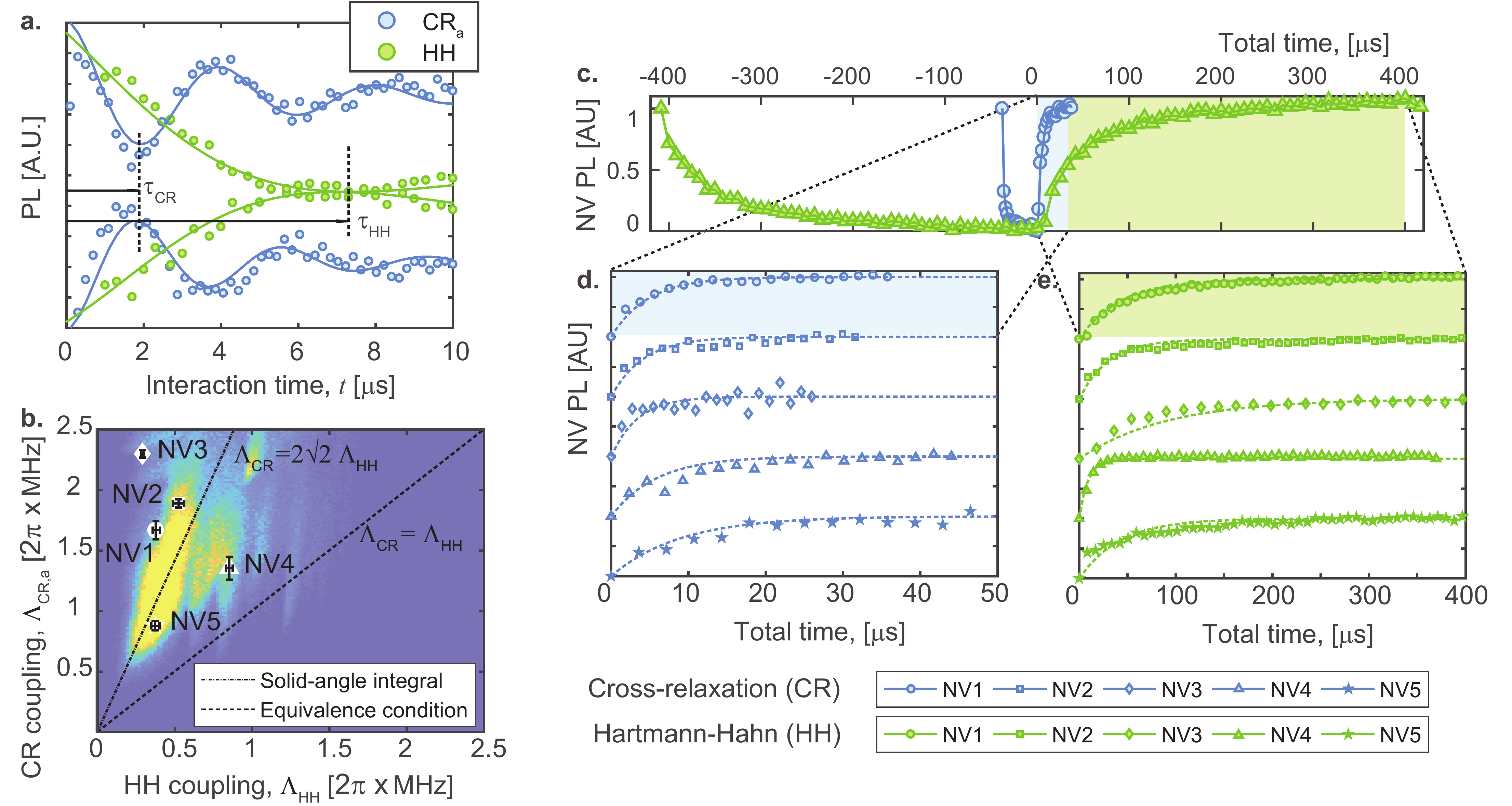}
  \caption{\textbf{Experimental study of native $^{13}$C polarisation.}
   \textbf{a.} Resonant longitudinal spin dynamics of NV1 under depolarisation (alternating CR/iCR (blue) and HH/iHH (green))  sequences. Fits are used to extract CR and HH hyperfine components as discussed in main text. \textbf{b.} Scatter plot of hyperfine couplings, $\Lambda_\mathrm{CR,a}$\,vs.\,$\Lambda_\mathrm{HH}$, extracted for NVs1-5 laid over their probability distribution (as taken from Fig.\,\ref{Fig_prob_dist}\,\textbf{e}.). Equivalence line ($\Lambda_\mathrm{CR,a}=\Lambda_\mathrm{HH}$, dashed) and average solid integrals ($\Lambda_\mathrm{CR,a}=2\sqrt{2}\Lambda_\mathrm{HH}$, dot-dashed) also shown. \textbf{c.} Polarisation and anti-polarisation buildup of NV1 under CR$_\mathrm{a}$ (circles) and HH (triangles) sequences. \textbf{d.} Zoomed plot of NV1-5 polarisations showing how polarisation timescales are related to CR hyperfine couplings. Shaded data corresponds to that of \textbf{c}. Theoretical curves (dashed) are constructed from extracted hyperfine coupling rates shown in \textbf{b}. \textbf{e.} As in \textbf{d}, but for HH polarisation sequence. } \label{Fig_3}
\end{figure*}

Using the same approach, we consider the joint probability distributions associated with the respective hyperfine distributions across the three schemes under consideration. Using the same approach as the 1-D distributions above, we bin the hyperfine realisations according to their location in $\Lambda_\mathrm{CR,a}-\Lambda_\mathrm{HH}$ (Fig.\,\ref{Fig_prob_dist}.\,\textbf{e}), $\Lambda_\mathrm{CR,a}-\Lambda_\mathrm{CR,b}$ (Fig.\,\ref{Fig_prob_dist}.\,\textbf{f}), and $\Lambda_\mathrm{CR,b}-\Lambda_\mathrm{HH}$ (Fig.\,\ref{Fig_prob_dist}.\,\textbf{g}) space. The white dashed lines denote the average relative scaling determined by integration of Eqs.\,\ref{LambdaRTheta} over the polar angle, $\Theta$; indicating that the coupling for CR$_\mathrm{a}$ is on average a factor of $2\sqrt{2}~2.8$ stronger than for HH.  From these distributions, we find that there is a 97\% chance that a given NV will exhibit a total CR$_\mathrm{a}$ hyperfine coupling greater that its HH coupling. On the other hand, the comparatively large overlap between the $\mathrm{Pr}_\mathrm{CR,b}$ and $\mathrm{Pr}_\mathrm{HH}$ results in a 66\% chance of CR$_\mathrm{b}$ coupling being the greater.

Having gained an understanding of the hyperfine coupling statistics, the following section is focused on experimentally examining the NV-$^{13}$C polarisation dynamics for a range of NV centres.

\subsection{Experimental results and discussion}\label{Subsec_13C_Results}
In order to experimentally compare the performances of the CR and HH schemes, five separate NV centres, denoted NV1-5, were studied from deep within an electronic grade diamond sample to reduce the likelihood of NV and $^{13}$C spins being perturbed by electron spins associated with nitrogen donor defects. This allowed us to exclusively examine the polarisation buildup dynamics of $^{13}$C nuclei for the two techniques without competing effects from electronic spin impurities within the diamond. As such, target spins in these samples exist in the $\Lambda>\Gamma_2$ regime, and thus exhibit coherent flip-flops due to their interaction with the NV centre under the relevant resonance matching condition.

The optimal interaction time, $\tau$, for each NV was determined by observation of the dynamics of the NV spin under the CR$_\mathrm{a}$ (HH) protocol as described above, by preparing the NV in its $|0\rangle$ ($|+\rangle$) state, and varying the NV-bath interaction time. As repeated application of this sequence would quickly polarise the target spins into a state that won't interact with the NV, bath polarisation was avoided by employing alternating polarisation/depolarisation sequences by using an additional RF $\pi$-pulse to prepare then NV in its $|-1\rangle$ ($|-\rangle$) state in every second application (Fig.\,\ref{Fig_3a}.\textbf{a}), thereby driving the target spins toward the opposite polarisation direction. As shown in Fig.\,\ref{Fig_3a}.\textbf{b}, by varying the interaction time, $\tau$, between pulses we are able to resolve the coherent flip-flop dynamics between the NV and its surrounding environment. Increasing the number of repetitions in the polarisation ($N_\mathrm{p}$) and depolarisation ($\mathrm{i}N_\mathrm{p}$) phases of the stationary pulse sequence ($N_\mathrm{p}=\mathrm{i}N_\mathrm{p}>1$) clearly shows that the flip-flop rate is preserved as nuclear polarisation increases and NV PL contrast reduces, as per Eqs.\,\ref{PolNoDephasing}\,and\,\ref{LambdaDef}, and as illustrated in Fig.\,\ref{Fig_2}\textbf{b}.

An example of the determination of the optimal interaction time for both schemes, $\tau_\mathrm{CR}$ and $\tau_\mathrm{HH}$, is shown in Fig.\,\ref{Fig_3}\,\textbf{a} for NV1.
By extracting the rate of coherent NV-$^{13}$C flip-flops for each NV, we compared their total CR and HH hyperfine couplings, $\Lambda_\mathrm{CR,a}$ and $\Lambda_\mathrm{HH}$, as shown in Fig.\,\ref{Fig_3}\,\textbf{b}. From this, we see that for all NVs studied, the CR$_\mathrm{a}$ coupling was greater than the HH coupling, as was expected from the discussions of the hyperfine coupling distributions above. To make this comparison clear, the scattered ($\Lambda_\mathrm{HH},\Lambda_\mathrm{CR,a}$) data was laid over the $P(\Lambda_\mathrm{HH},\Lambda_\mathrm{CR,a})$ distribution from Fig.\,\ref{Fig_prob_dist}.\,\textbf{e}.

In order to confirm the relationship between coupling strengths and nuclear spin polarisation rate, we studied the rate of polarisation buildup under CR$_\mathrm{a}$ and HH schemes. Any unknown remnant or statistical polarisation was first removed by polarising the target spins anti-parallel to the background field by using 20\,iCR (50\,iHH) inverse pulses with interaction times of $\tau_\mathrm{CR}$ ($\tau_\mathrm{HH}$) defined by the respective hyperfine couplings in Fig.\,\ref{Fig_3}\,\textbf{b}. This was followed by 20\,CR$_\mathrm{a}$ (50\,HH) pulses to polarise the target spins parallel to the background field. An example of the full time dependence of this approach is shown in Fig.\,\ref{Fig_3}\,\textbf{c} for NV1. Zoomed plots of the respective polarisation buildup periods (shaded regions in Fig.\,\ref{Fig_3}\,\textbf{c}) for CR and HH schemes across the five NVs studied are shown in Figs.\,\ref{Fig_3}\,\textbf{d}\,\&\,\textbf{e}, respectively, together with the theoretically expected polarisation behaviour.

The theoretical curves were constructed by summing over all spins $j$ in Eq.\,\ref{DEDisc}, and recognising that, under the assumption that nuclear spin relaxation $(\Gamma_1)$ and spin diffusion rates $(D)$ processes do not contribute on timescales of $t\sim 1/\Lambda$, the polarisation gained by the $^{13}$C target spins is equal to that lost by the NV,
\begin{eqnarray}
  P_\mathrm{NV}(t) &=& 1-\exp\left(-\frac{1}{\pi}\Lambda_\mathrm{total}t\right);
\end{eqnarray} and evaluations of the total polarisation buildup rates $(\Lambda_\mathrm{total})$ proceed via the summation defined in Eq.\,\ref{LambdaDef},
which were taken directly from the total hyperfine couplings for each NV (data extracted as per Fig\,\ref{Fig_3} \textbf{a.} and mapped in Fig\,\ref{Fig_3} \textbf{b.}). Comparison of the theoretical and experimental data in Figs\,\ref{Fig_3} \textbf{c.}\,\&\,\textbf{d.} show good agreement, and confirming the polarisation rate of CR$_\mathrm{a}$ is significantly faster than that of HH.

Environmental polarisation is inferred by measuring the likelihood of the NV to tunnel out of its initial spin state: if the bath is in a polarised state with which the NV spin cannot interact, the NV will remain in its initial state. Further polarisation of the target ensemble must therefore rely on dipole-mediated diffusion for polarisation transport to regions beyond the reach of the hyperfine field within the relaxation time of the target spins. For the $^{13}$C case studied here, such processes occur on much longer timescales than $\tau$, and are hence not considered in this section. These processes were treated in detail in Ref.\,\onlinecite{Hal17}, and are critical to the study of target ensembles external to the diamond undertaken in the following section.

\begin{figure*}
  \centering
  \includegraphics[width=\textwidth]{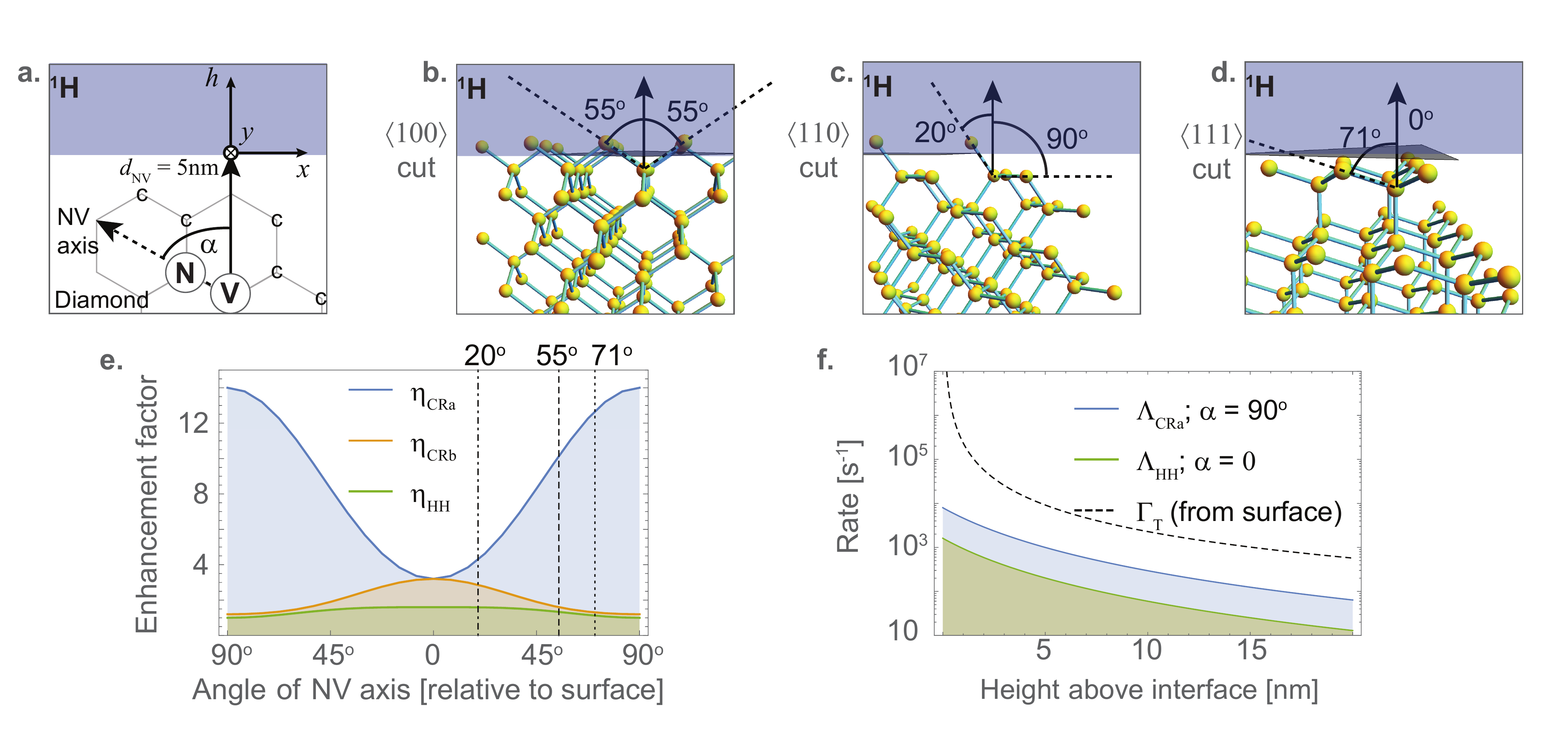}
  \caption{\textbf{Dependence of NV-target coupling on surface orientation.} \textbf{a.} Schematic of NV axis orientation, $\alpha$, with respect to diamond-target interface. \textbf{b-d.} Schematics of three common cut orientations seen in single-crystal diamond ($\langle100\rangle$, $\langle110\rangle$, and $\langle111\rangle$) together with relative NV/surface orientations. \textbf{e.} Ratio of total hyperfine couplings for CR$_\mathrm{a}$, CR$_\mathrm{b}$, and HH cases for surface orientations $\left[-90^\circ,90^\circ\right]$ relative to HH$\left(90^\circ\right)$ case. Respective maxima occur for NVs oriented at $90^\circ,\,0^\circ\,0^\circ$ relative to the surface. \textbf{f.} Comparison of the hyperfine ($\Lambda$) and dephasing ($\Gamma_\mathrm{T}$) fields felt by the target spins vs height above the diamond surface for a 5nm deep NV, showing that NV-target interactions exist in the strongly dephased regime in all cases.
  }\label{Fig_4}
\end{figure*}

\section{Prospects for hyperpolarisation of external macroscopic nuclear spin ensembles}\label{External_prospects}
We now study how the choice of polarisation protocol will affect the polarisation extent (total number of spins polarised) of a target nuclear spin ensemble external to the diamond crystal, as depicted in Fig.\,\ref{Fig_4}.\textbf{a}. For this study, we consider the proton spins in a semi-infinite slab of poly(methyl methacrylate) (PMMA) as our example ensemble, as this system has been well studied in the past in the context of both NV-based NMR\,\cite{Mam13,Stu13,Woo17} and DNP\cite{Hal17} protocols. The PMMA parameters of physical relevance to this study are a proton density of $n_\mathrm{p}=56\,\mathrm{nm}^{-3}$, and an average spin lattice relaxation time of $T^\mathrm{T}_1=1$\,s\,\cite{Hat76}; and we focus our discussion on NV centres residing approximately 5nm below the diamond-PMMA interface, as this provides an acceptable compromise between minimising NV decoherence\,\cite{San19}, and maximising NV-target couplings.

In order to compare the relative polarisation rates and resulting yields across pulse schemes, it is necessary to first understand the effects and physical origins of the hyperfine, dephasing, and motional processes associated with this system geometry. To this end, we have organised this section as follows. We begin by describing the adaptation of the dynamical system described in Eq.\ref{DEDisc} to a continuum model to accommodate the increased number of target spins interacting with the NV due to their motion. We then separately describe the physical and quantum coherence properties of the NV array and the target nuclei by accounting for the new geometry, as well as dephasing and diffusion effects.
Finally, we use this description to compare CR$_\mathrm{a}$, HH, and PP protocols under realistic experimental conditions over a wide range of target diffusion rates, and conclude with a discussion of both idealised and practical performance limits for each protocol and the conditions under which these limits could be realised.

\subsection{Continuum description}\label{Sec_Analysis_cont_desc}
Accurate modelling of these systems typically requires consideration of regions of some ~200\,nm or larger in size. This equates to $\sim10^{10}$ proton spins and $10^{20}$ couplings in a typical organic target sample, thereby rendering its description beyond the reach of the discrete dynamical model given in Eq.\,\ref{dynamical_system}. This is further complicated if the lengthscales of target motion over the timescale of $T_1^\mathrm{T}$ exceeds the region's size. The relatively high target densities ($\sim50\,\mathrm{nm^{-3}}$) compared with NV-target standoffs ($5-10\,\mathrm{nm}$) however allows for the mapping of this system to a continuum description, in which we are concerned with the polarisation of a fixed infinitesimal volume at location $\mathbf{R}$, of which individual spin may move in or out; rather than tracking the trajectories and polarisations of individual particles. In Ref.\,\onlinecite{Hal17}, it was shown for the CR case, such a system may be modelled via a convection-diffusion equation with spatially dependent cooling rate
\begin{eqnarray}
  u_j &\mapsto& u(\mathbf{R}).
\end{eqnarray}
Given the theoretical correspondence between CR and other schemes demonstrated in Section\,\ref{BulkC13Sec}, this description holds for all such schemes, up to a respective definition of $u(\mathbf{R})$,
\begin{eqnarray}
  \frac{\partial}{\partial t} P(\mathbf{R},t) &=& -u(\mathbf{R})\left[P(\mathbf{R},t)+\frac{1}{2}\right] -\Gamma_1P(\mathbf{R},t)\nonumber\\
  &&+ D\nabla^2P(\mathbf{R},t),\label{DECont}
\end{eqnarray}
where $D=D_\mathrm{hf}+D_\mathrm{dp}+D_\mathrm{sp}$ is the total effective diffusion constant due to processes associated with hyperfine\,\,(hf) mediated, nuclear dipole-dipole\,\,(dp) mediated, and spatial\,\,(sp) diffusion respectively.

By virtue of such target mixing, we would expect the existence of an optimal diffusion rate that maximises the polarisation yield. As target motion begins to become appreciable on the timescales of nuclear relaxation, we expect the total polarisation yield to increase owing to polarised targets being transported away from the hyperfine field and replenishing the saturated region with unpolarised spins. However, as these dwell times approach the timescales of the hyperfine interaction, we expect a dramatic reduction in the cooling rate, and a corresponding reduction in polarisation yield. These effects will be of critical importance in comparing the performance of each scheme, and will be detailed explicitly below in Sections\,\ref{Sec_ext_hyperfine}\,\&\,\ref{Sec_ext_dephasing}.

\subsection{Analysis of NV-target interaction.}
\label{Sec_ext_hyperfine}
 Critically, as suggested by the hyperfine lobe geometries in Fig.\,\ref{Fig_1}\,\textbf{d.}, the relative orientations of the NV axis and the diamond surface (Fig\,\ref{Fig_4}\,\textbf{a.}) allow for a large variation in the overall hyperfine cooling rate, as these determine the effective number of target spins residing in the hyperfine lobes of the scheme under consideration.
 In this section we determine the optimal surface orientation and resulting spatial dependencies of the hyperfine field for each scheme, and discuss how this may be realised in the context of common surface cut orientations (Figs\,\ref{Fig_4}\,\textbf{b-d.}). Organic $^1$H polarisation targets such as the PMMA discussed here and in ref.\,\onlinecite{Abr14}, and demonstrated in ref.\,\onlinecite{Hal17}, typically exhibit nearest neighbour separations that are 1-2 orders of magnitude smaller than the depth of the NV below the surface, allowing the ensemble to be treated as a continuum. As such, we employ the following total squared hyperfine field integral to determine the optimal orientation of each scheme:
\begin{eqnarray}
  \Lambda_\mathrm{Total}^2(\alpha) &=&  \int_\mathrm{ext}n_\mathrm{p}(\mathbf{R_\alpha})\Lambda^2(\mathbf{R_\alpha})\,\mathrm{d}^3\mathbf{R_\alpha},\label{Lambda_ext}
\end{eqnarray} where $n_\mathrm{p}$ is the proton density in the target region, and $\int_\mathrm{ext} (\cdots)\,\mathrm{d}^3\mathbf{R_\alpha}$ refers to integration over the target region external to the diamond crystal for a surface orientation of $\alpha$. The reasons for comparing the integral of the squared hyperfine field are due to the cooling rate in the high-dephasing regime being proportional to this quantity, as will be justified below in Section\,\ref{Sec_ext_dephasing}.

To compare the protocol-specific hyperfine rates for surface orientations of $-90^\circ\leq\alpha\leq90^\circ$ relative to the NV axis, we plot the relative hyperfine enhancement factor given by
\begin{eqnarray}
  \eta(\alpha) &=& \frac{\Lambda_\mathrm{Total}^2(\alpha)}{\Lambda_\mathrm{Total}^2(\mathrm{HH};\alpha=90^\circ)},
\end{eqnarray}where $\Lambda_\mathrm{Total}^2(\mathrm{HH};\alpha=90^\circ)$ refers to the HH case with the NV axis oriented perpendicular to the surface normal. From the plots of $\eta_\mathrm{CRa}$, $\eta_\mathrm{CRb}$, and $\eta_\mathrm{HH}$ in Fig.\,\ref{Fig_4}\,\textbf{e} we see that the total squared hyperfine coupling for the CR$_\mathrm{a}$ scheme is greater than that of the other two for all possible orientations, with the orientation-optimised enhancement being an order of magnitude greater between the CR$_\mathrm{a}$ and HH cases.
This is a direct consequence of the hyperfine lobe structure associated with each scheme, and has implications for the relative number of spins that may be polarised under each scheme once the hyperfine cooling has reached equilibrium with spin diffusion and spin-lattice relaxation processes.

From Fig.\,\ref{Fig_4}\,\textbf{e}, we see that the maximal CR$_\mathrm{a}$ cooling rate will be realised when the NV axis lies parallel to the surface plane (perpendicular to the surface normal), whereas the opposite is true for CR$_\mathrm{b}$ and HH cases. The 90$^\circ$ case may be realised by aligning the external field along one of the $\langle1\bar{1}\bar{1}\rangle$ or $\langle\bar{1}1\bar{1}\rangle$ families of NV centres in a crystal with the diamond-target interface cut perpendicular to the $\langle110\rangle$ direction (Fig.\,\ref{Fig_4}\,\textbf{c}); whereas the 0$^\circ$ cases may be realised using the $\langle111\rangle$ family of NV axes in a $\langle111\rangle$ cut crystal (Fig.\,\ref{Fig_4}\,\textbf{d}). The most common commercially available surface cut orientation of $\langle100\rangle$ ($\alpha=55^\circ$) is also shown in Fig.\,\ref{Fig_4}\,\textbf{b}.

Evaluating Eq.\,\ref{Lambda_ext} under these respectively optimised orientations using the hyperfine fields given in Eqs.\,\ref{LambdaRTheta} \,\&\,\ref{LambdaPulses}, the squared hyperfine couplings between the NV at a depth $d$ below the surface and an NV-proton separation of $\mathbf{R}=(x,y,d+h)$ (Fig\,\ref{Fig_4}\,\textbf{a.}) are:
\begin{eqnarray}
  \Lambda^2_\mathrm{CRa}(\mathbf{R}) &=& \frac{9 a^2 \left((d+h)^2-y^2\right)^2}{2 \left((d+h)^2+x^2+y^2\right)^5},\nonumber\\
  \Lambda^2_\mathrm{HH}(\mathbf{R}) &=& \frac{9 a^2 (d+h)^2 \left(x^2+y^2\right)}{4 \left((d+h)^2+x^2+y^2\right)^5},\nonumber\\
  \Lambda^2_\mathrm{PP}(\mathbf{R}) &=& \frac{4}{9\pi^2}\left(6+4\sqrt{2}\right)\Lambda^2_\mathrm{HH}(\mathbf{R}).\,\,\,\,\,\,\,
  \label{Eq_ext_Lambda}
\end{eqnarray}

In the case of pulsed protocols, there is an additional cooling reduction that arises when target nuclei are not in contact for the duration of a pulse cycle. Specifically, these protocols are engineered using time-dependent Hamiltonians to mimic an electron-nuclear flip-flop operation, but only do so if the target is coupled to the NV for the duration of the pulse cycle; if not, the interaction will, on average, be predominantly suppressed. The result of this is that the effective hyperfine coupling for pulsed (p) schemes is attenuated via
\begin{eqnarray}
  \Lambda^2_\mathrm{p} &\mapsto&  \Lambda^2_\mathrm{p}\frac{\tau_\mathrm{d}^2}{\tau_\mathrm{d}^2+\tau_\mathrm{p}^2},
  \label{Eq_Lambda_damping}
\end{eqnarray}where $\tau_\mathrm{d}$ is the target's `dwell time' for which it is coupled to the NV. The interaction cutoff time, $\tau_\mathrm{p}$, depends on the pulse sequence chosen but is of the order of its pulse spacing, as is typically defined by the target's Larmor frequency, $\omega_\mathrm{T}$. 
In the case of PulsePol, we have $\tau_\mathrm{p} = 3\pi/\omega_\mathrm{T}$, and the pulse spacing would be $\tau/4$, with the full dependence thus given by
\begin{eqnarray}
  \Lambda^2_\mathrm{PP}(\mathbf{R}) &=& \frac{6+4\sqrt{2}}{\pi^2}\frac{ a^2 (d+h)^2 \left(x^2+y^2\right)}{ \left((d+h)^2+x^2+y^2\right)^5}\frac{\tau_\mathrm{d}^2}{\tau_\mathrm{d}^2+\tau_\mathrm{p}^2}.\,\,\,\,\,\,\,\,\,
  \label{Eq_ext_Lambda_2}
\end{eqnarray}

For the non pulsed protocols (CR and HH), this effect is not observed as the nuclei are coupled to the NV in a `passive` or time-independent manner. In this case, the effects of target motion in the region surrounding some separation $\mathbf{R}$ effectively present as an additional source of dephasing in that region. This will be explored in the following subsection.

Having determined the optimal surface orientations to maximise the hyperfine field for each scheme, and their associated spatial dependencies, we now turn to the elucidation of their respective cooling rates for implementation of the polarisation transport equation (Eq.\,\ref{DECont}).

\subsection{Hyperfine cooling rates under strong dephasing}
\label{Sec_ext_dephasing}
As discussed in the previous section, a detailed understanding of the polarisation of external spin requires an understanding of the way in which target motion and dephasing affect the polarisation transfer and transport processes. These processes arise from both target motion and non-target paramagnetic impurities coupling to both the NV and target spins. In what follows we extend the theoretical framework of Section\,\ref{BulkC13Sec} to include these effects.

In most applications of practical importance, additional paramagnetic impurities exist in close proximity to the polarisation source and target. In the case of NV centres in diamond, electronic spin impurities often arise from substitutional P1 nitrogen donor spins\cite{Han08}, surface electron spins\cite{Gri14,Ros14,Mye14,Lua15,Rom15}, and additional damage created during nitrogen implantation processes\cite{Leh16,Tet18,Hea20}. These give rise to strong additional sources of magnetic noise, which act to dephase the NV-target spin interaction, and ultimately slow the polarisation transfer rate.

Whilst the dephasing rate of the NV will vary between protocols, consideration of the nuclear dephasing in the following section shows that it exceeds the hyperfine coupling in all cases of practical relevance. Consequently, all nuclei in this system will necessarily exist in the strong-dephasing regime ($\Gamma(\mathbf{R})\gg\Lambda(\mathbf{R})$) regime, implying the cooling field as described by Eq.\,\ref{Eq_cool_rate_coherent} and used for the $^{13}$C nuclei studied in the previous section, does not apply here. As such, we return to Eq.\,\ref{Eq_Lindblad}, but instead seek solution under the $\Gamma_2 = \Gamma_\mathrm{NV} + \Gamma_\mathrm{T}>\Lambda$ regime.
Here we find the polarisation behaviour modified from the flip-flop behaviour described in Eq.\,\ref{PolNoDephasing} to and exponential approach to equilibrium,
\begin{eqnarray}
  P(t)&=& P(0)+\frac{1}{2} \left(P(0)+\frac{1}{2}\right) \left[\exp\left({-\frac{ \Lambda^2}{\Gamma_2}}t\right)-1\right]\label{PolWithDephasing},\,\,\,\,\,\,\,\,\,
\end{eqnarray} where both $\Lambda$ and $\Gamma$ depend on the specific scheme in question.
Critically, this equation shows that the polarisation rate is suppressed as the dephasing rate is increased beyond the total hyperfine coupling. Thus, the task of determining the effective cooling rates requires a precise understanding of the sources of dephasing present.

For the CR case, the dephasing rate between the NV \ket0 and \ket{-1} states is determined from the transverse spin relaxation rate during a Ramsey experiment, $\Gamma_\mathrm{CR}=1/T_2^*$, which depends primarily on the surrounding paramagnetic environment. In the HH case, the dephasing between the \ket+ and \ket- states under microwave driving (often referred to as $1/T_{1,\rho}$) is determined via the longitudinal relaxation rate during a Rabi experiment. In the pulsed cases considered here, the dephasing rate is equivalent to that measured under an XY dynamical decoupling sequence of commensurate order $n$; often denoted $1/T_2^{XY,n}$.

It is also important to note that whilst the decoherence rate of the NV spin can be reduced using HH or pulsed DNP protocols, these protocols do not improve the coherence properties of the target nuclear spins. If the latter makes the dominant contribution to the decoherence of the NV-target interaction, then there will be no reduction in the overall $\Gamma_2$ to be gained from the implementation of pulsed control of the NV alone. The decoherence of NV spins depend on the distributions of all paramagnetic impurities present, their interactions with each other and the NV, and the static and RF field strengths\,\cite{Dob09,Hal14,Lan18}; the effects of which will be summarised below.

In any case, under the $\Gamma_2 = \Gamma_\mathrm{NV} + \Gamma_\mathrm{T}>\Lambda$ regime, we identify the cooling rate for use in Eq.\,\ref{DECont} to be
\begin{eqnarray}
  u(\mathbf{R}) &=& \frac{\Lambda^2(\mathbf{R})}{\Gamma_2(\mathbf{R})}.
  \label{Eq_cool_rate_incoherent}
\end{eqnarray} Having done so, we now quantify the dephasing contributions of the electron and nuclear spins separately, as well as the dephasing effects of target motion. These will be combined with the hyperfine discussions above to evaluate the effective cooling rates for each scheme in the following section.

\subsubsection{Electron spin dephasing}
For the foreseeable future, any diamond material to be used for ensemble NV-based DNP of external nuclei will contain a large proportion of non-NV paramagnetic defects that act to dephase the electron ensemble on $\sim$0.1-1\,MHz timescales, which are faster than the $\sim$10-100\,KHz rates at which polarisation may be transferred to the nuclear ensemble.
Therefore, any practical external nuclear polarisation process will necessarily exist in an $\Lambda\ll\Gamma_\mathrm{NV}$ regime.
For shallow NVs, typical reported N-to-NV conversion rates are just a few percent\,\cite{Fav17,Hea20}. Of those, only a quarter will have the necessary alignment to be utilised for polarisation of the nuclei, meaning that the vast majority of the electron spins associated with the nitrogen present will act only to dephase the tiny fraction ($<~1$\%) utilised for DNP. This is worsened in cases when the electron ensemble array is created via implantation of nitrogen ions, as presently necessitated by the requirement for a thin electron array to exist within $\sim10\,$nm of the diamond-target interface. This results in additional damage to the lattice in the form of vacancy chains, and results in further electron spin noise\,\cite{Tet18}.

As such, for the purposes of this discussion, we adopt the current state of the art NV ensemble properties used for NV based NMR sensing, as these exhibit the optimal conditions for maximising NV density and minimising competing NV decoherence effects:
\begin{eqnarray}
  \Gamma_\mathrm{CR_a}^\mathrm{NV} &=& 1\,\upmu\mathrm{s}, \nonumber\\
  \Gamma_\mathrm{HH}^\mathrm{NV} &=& 2\,\upmu\mathrm{s}, \nonumber\\
  \Gamma_\mathrm{PP}^\mathrm{NV} &=& 20\,\upmu\mathrm{s}.
  \label{Eq_ext_gamma_NV}
\end{eqnarray}

\subsubsection{Nuclear spin dephasing}
Whereas the implanted nitrogen layer is the dominant source of dephasing for the NV spin, the spin-spin dephasing of the target nuclei is predominantly the result of coupling to electron spins on the diamond surface.
This effect is primarily determined by the surface electron spin density, and a multitude of NV measurements have determined that these spins baths present with a g-factor of 2, and aerial densities of $\sigma_\mathrm{surf}\sim0.01-0.5\,\mathrm{nm}^{-2}$ (Refs.\,\onlinecite{Gri14,Ros14,Mye14,Lua15,Rom15}) depending on the nature of the surface preparation\,\cite{San19}. Fluorene terminated surfaces have been reported to exhibit the lowest densities of around $\sigma_\mathrm{surf}\sim0.01\,\mathrm{nm}^{-2}$ (Ref.\,\onlinecite{Ros14}), nanodiamonds exhibit the highest densities of around $\sigma_\mathrm{surf}\sim0.5\,\mathrm{nm}^{-2}$ (Ref.\,\onlinecite{Gri14}), whereas the implanted samples considered here result in densities of approximately $\sigma_\mathrm{surf}\sim0.1\,\mathrm{nm}^{-2}$. We therefore take the latter as the value for this analysis.

The decoherence rate imparted by the surface spins on nuclei at a height $h$ above the surface is given by
\begin{eqnarray}
  \Gamma_\mathrm{T} &=& \frac{\sqrt{3 \pi a  \sigma_\mathrm{surf}} }{16 h^2}\sqrt{  20 \cos (2 \alpha )+3 \cos (4 \alpha )+41},\,\,\,\,\,\,\,
  \label{Eq_ext_gamma_T}
\end{eqnarray}where $\alpha$ is the polar angle between the direction of the external field (as aligned with the NV axis) and the surface normal, thus defining the quantisation axis of the target and surface spins.

\subsubsection{Motion of target nuclei}
\label{Sec_Analysis_motion}
In Section\,\ref{BulkC13Sec} we noted the importance of diffusion and convection in avoiding polarisation saturation by transporting polarisation to regions of the target beyond the reach of the hyperfine field. These processes, however, can also have a deleterious effect on the cooling rate if the target's motion causes its hyperfine interaction time to be reduced below the optimal hyperfine interaction time. For the CR and Novel cases, this can be understood as follows: i) the NV spin entangles with target nuclei in a given region, ii) the target nuclei are transported from this region to one out of reach of the NV hyperfine field, resulting in a loss of quantum phase information associated with the NV and those nuclei, iii) the region is replenished with nuclei having no phase correlation with the NV. This manifests as an additional component to the dephasing rate,
\begin{eqnarray}
  \Gamma_2(\mathbf{R}) &=& \Gamma_2^{\mathrm{NV}}+\Gamma_2^{\mathrm{T}}(\mathbf{R})+\Gamma_2^{\mathrm{diff}}(\mathbf{R}),\label{Eq_total_dephasing}
\end{eqnarray}where
\begin{eqnarray}
  \Gamma_2^{\mathrm{diff}}(\mathbf{R}) &=& \frac{8}{\tau_\mathrm{d}(\mathbf{R})},\label{Eq_diff_from_tau_d}
\end{eqnarray} and $\tau_\mathrm{d}(\mathbf{R})$ is the average amount of time nuclei at position $\mathbf{R}$ are in contact with the NV as a result of their motion. 
The position dependence arises due to nearby targets having a shorter effective dwell time than those further from the NV.

For the present case we assume that there is no externally influenced convective motion of the target protons, and that polarisation transport is dictated by physical or dipole-mediated spin-spin diffusion processes only. We use Eq.\,\ref{Eq_diff_from_tau_d} to relate the resulting contribution to the dephasing rate to the effective dwell time of the nuclei in the vicinity of the NV. 
This dwell time is in turn linked to the total diffusion constant, $D$ via
\begin{eqnarray}
  \tau_\mathrm{d} &=& \frac{\left(h+d\right)^2}{6 D}.
\end{eqnarray}

Following the discussion in Section\,\ref{Sec_Analysis_cont_desc}, the total diffusion coefficient, $D$, arises from the combined processes of physical diffusion and spin diffusion,
\begin{eqnarray}
  D &=& D_\mathrm{dp} + D_\mathrm{sp} ,
\end{eqnarray}where we have ignored the effects of hyperfine-mediated diffusion (as detailed in Eq.\,\ref{DEDisc}) on the grounds that we expect nuclei of equal hyperfine coupling to have equal degrees of polarisation after initialising the target ensemble in a mixed state. Whilst this may not be strictly true in regions very close to the diamond-target interface due to large dephasing gradients, we do not expect this to affect the steady state polarisations with which we are concerned in this section.

The dipole-mediated diffusion constant, $D_\mathrm{dp}$, of the target could, in principle, depend explicitly on the spatial diffusion constant, $D_\mathrm{sp}$, via the resulting motional narrowing of dipole interactions amongst target constituents. For the system considered here however, this effect is always weaker than direct dipole coupling or spatial diffusion, and is thus ignored in what follows. 
We therefore define the dipole mediated diffusion constant in terms of the average nearest-neighbour dipole flipping rate ($f_\mathrm{dip}$) and proton separation ($ r_\mathrm{nn}$) via
\begin{eqnarray}
  D_\mathrm{dp} &=& \frac{1}{6}f_\mathrm{dip} r_\mathrm{nn}^2\nonumber\\
  &\sim&\frac{\gamma_\mathrm{T} ^2 \mu _0 \sqrt[3]{n} \hbar }{24 \pi } ,\nonumber\\
  &\approx& 455\,\mathrm{nm^2\,s^{-1}},
\end{eqnarray}
with $D_\mathrm{sp}$ left as a parameter for us to vary as part of the scheme comparison below.

\begin{figure*}
  \centering
  \includegraphics[width=\textwidth]{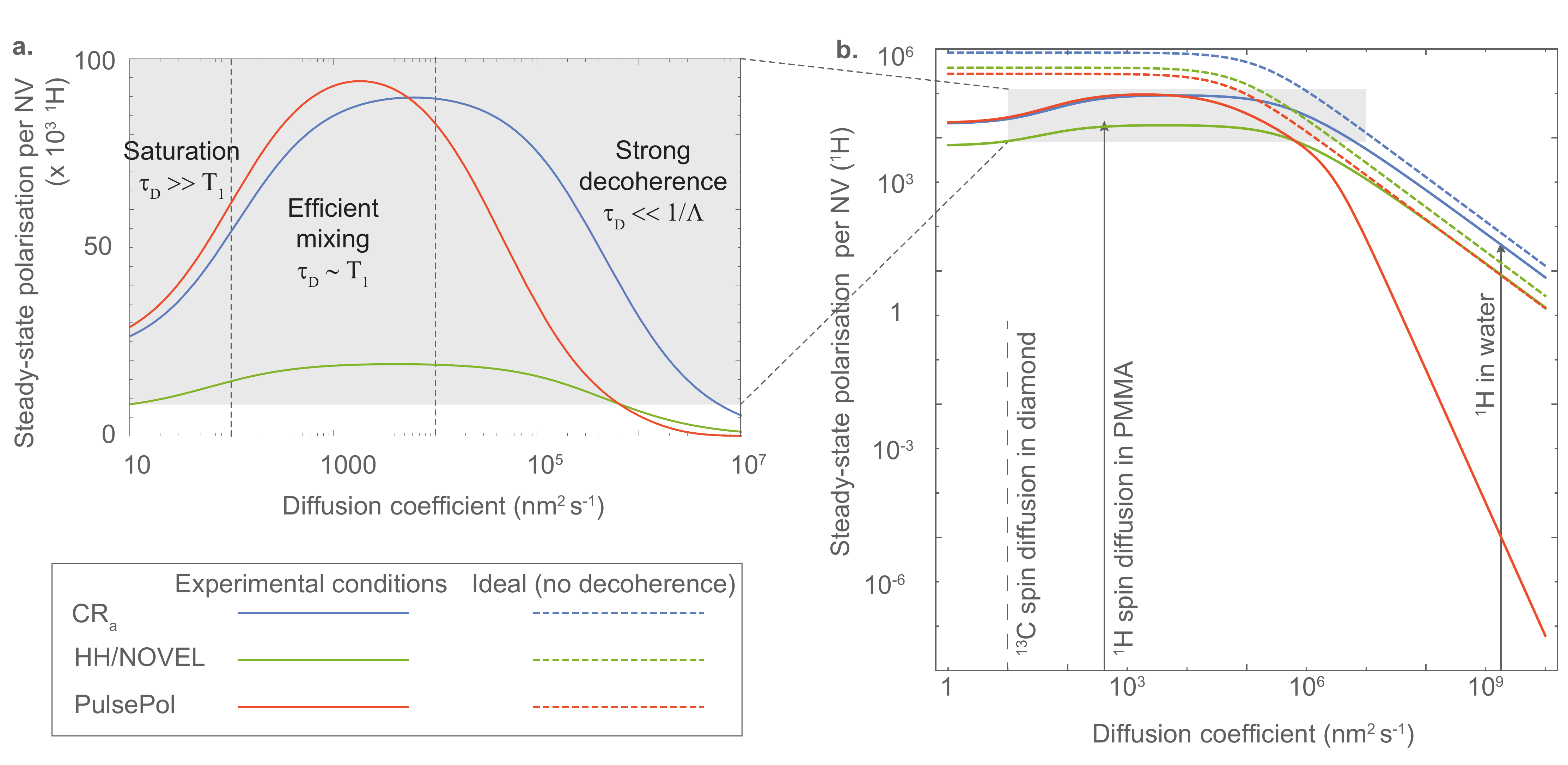}
  \caption{\textbf{Comparison of NV-DNP polarisation yields. a.} Log-linear plot showing the scaling of the total external $^1$H polarisation achievable with a single NV using CR$_\mathrm{a}$ (blue), NOVEL (green), and PulsePol (red) NV-DNP schemes.
  \textbf{{b.}} Zoomed-out plot of total polarisation vs diffusion coefficient for the same system as in \textbf{{a}} showing the full scaling of each scheme under experimental conditions (solid curved). Dashed curves show the expected yields under the idealised low decoherence regime ($\Gamma_2\ll\Lambda$). Diffusion coefficients of various systems are indicated for the sake of comparison.
  Assumed parameter values: $T_2^* = 1\,\upmu\mathrm{s}$, $T_{1,\rho} = 2\,\upmu\mathrm{s}$, $T_2 = 20\,\upmu\mathrm{s}$, $n_p=56\,\mathrm{nm^{-3}}$, $\sigma_e = 0.1\,\mathrm{nm^{-2}}$, $d_\mathrm{NV}=5\,\mathrm{nm}$, $T_1^\mathrm{T}=1\,\mathrm{s}$.
  } \label{Fig_sspol_comparison}
\end{figure*}

\subsection{Comparison of polarisation yields and discussion}
In this section, we will use the polarisation transport equation, Eq.\,\ref{DECont}, to determine the polarisation yields of each scheme under both realistic and ideal (decoherence-free) conditions. Having determined both the strength and coherence properties of the NV-target hyperfine interaction, we first determine the cooling rates of the target ensemble.

Comparison of the hyperfine couplings (Eq.\,\ref{Eq_ext_Lambda}) and dephasing rates (Eq.\,\ref{Eq_ext_gamma_NV}\,\&\,\ref{Eq_ext_gamma_T}) confirms that the NV-target system exists well within the high-decoherence ($\Gamma_2\gg\Lambda$) regime in all possible cases, as shown in Fig.\,\ref{Fig_4}.\textbf{f.}. In this regime, using Eq.\,\ref{Eq_cool_rate_incoherent}, the cooling rate field felt by target nuclei at position $\mathbf{R}$ is given by
\begin{eqnarray}
  u(\mathbf{R}) &=& \frac{\Lambda^2(\mathbf{R})}{\Gamma_2(\mathbf{R})}.
\end{eqnarray}
Using this result to determine the solution of Eq.\,\ref{DECont} for the polarisation field, $P(\mathbf{R},t)$ we may evaluate the total steady-state polarisation yield via
\begin{eqnarray}
  P_\infty &\equiv& \lim_{t\rightarrow\infty} \int_\mathrm{ext} n P(\mathbf{R},t)\,\mathrm{d}^3\mathbf{R},
\end{eqnarray}where again $\int_\mathrm{ext} (\cdots)\,\mathrm{d}^3\mathbf{R}$ refers to integration over the target ensemble external to the diamond.
This system was solved with Mathematica (NDSolve) using no-flux $\left(\nabla_\mathbf{R} P\cdot\mathbf{n}_\mathrm{boundary} = 0\right)$ boundary conditions. For diffusion constants below $10^4\,\mathrm{nm^2\,s^{-1}}$, cubic domains of sidelength $L = 1\,\upmu\mathrm{m}$ centred above the NV were employed. For larger diffusion coefficients, sidelengths of $L = 10\times\sqrt{6DT_1^\mathrm{T}}$ were employed in order to account for increasingly rapid polarisation transport away from the NV centre.

In Fig.\,\ref{Fig_sspol_comparison} we directly compare the expected polarisation yields for CR, NOVEL, and PulsePol schemes over the range of possible target diffusion rates under the realistic experimental conditions outlined in this section. Fig.\,\ref{Fig_sspol_comparison}.\,\textbf{a.} shows that in all cases, the polarisation exhibits an initial increase with increasing diffusion. This is due to the fact that when diffusion is not significant, the polarisation in the vicinity of an NV becomes saturated due to the localised nature of the hyperfine coupling. As diffusion increases, this saturation is able to be transported to regions beyond the reach of the hyperfine field before saturation occurs. The region surrounding the NV is then replenished with less polarised spins, which are subsequently polarised under the action of the cooing field.

Consideration of diffusion timescales $\tau_\mathrm{d}= (h+d)^2/6D$ confirms that this increase occurs when $\tau_\mathrm{d}\sim T_1^\mathrm{T}$ (leftmost dashed vertical line); that is, when diffusion-mediated mixing becomes appreciable on the lifetime of the imparted polarisation. Such diffusion rates are well-matched to dipole-mediated spin diffusion within solid-state or extremely viscous targets; which for $^1$H spins are typically of order 500\,$\mathrm{nm^2\,s^{-1}}$. As such, any appreciable spatial diffusion would likely drive this system outside of this regime, implying that organic protons in solid or extremely viscous states represent near-optimal motion for this application.

Under both saturated and efficient mixing regimes, we can see that the PulsePol scheme yields the greatest degree of polarisation, despite the significantly lower hyperfine coupling compared with other schemes. This is partially due to the roughly 20-fold improvement in NV coherence afforded by PulsePol offsetting the hyperfine advantage afforded by CR; however it should also be noted that,
 owing to electronic spins on the diamond surface, nuclei residing within the first few nanometres of the surface will exhibit stronger decoherence than the NV, and this decoherence is independent of the NV-DNP scheme used. Critically, this also suggests that if magnetic impurities could be removed from within or on the diamond, we would expect much greater performance under all schemes; the greatest of which would occur under CR. This is explored in more detail below.

As diffusion rates increase beyond the efficient mixing regime, dwell times of nuclei within the vicinity of the NV will become commensurate with, or shorter than the desired NV-target interaction time. Under this scenario, rapid replenishing of targets within this region constitutes an effective dephasing of the NV spin's interaction with proximate target nuclei, which make the overwhelmingly dominant contribution to the total decoherence rate, $\Gamma_2\sim 48D/d^2$ (Eq.\,\ref{Eq_total_dephasing}). This scaling is consistent across all schemes, and has the effect of significantly undermining any gains in NV coherence offered by PulsePol and Novel. As such, the sole point of comparison between schemes as diffusion increases beyond that due to dipole-dipole interactions is their relative hyperfine strengths. Moreover, as the effective hyperfine strength of PulsePol also strongly depends on this dwell time (as per the discussion around Eq.\,\ref{Eq_Lambda_damping}), we expect CR to perform dramatically better than other schemes when target diffusion coefficients exceed approximately $5\times10^4\,\mathrm{nm^2\,s^{-1}}$; albeit with significant reductions in performance across all schemes. This can be clearly seen in the strong decoherence region in Fig.\,\ref{Fig_sspol_comparison}.\,\textbf{a.}.

The scalings of each scheme's performance against diffusion can be seen more clearly in the zoomed out plot in Fig.\,\ref{Fig_sspol_comparison}.\,\textbf{b.}, which expands upon the ranges considered in \textbf{a}. From the logarithmic axes, we can see the rapid drop fall-off of polarisation extent for all schemes. For diffusion constants below approximately $D\sim100\,\mathrm{nm\,s^{-1}}$, as is the case for many solid-state NMR targets, the total polarisation appears to scale linearly with increasing diffusion. This scaling is 
plotted in Fig.\,\ref{Fig_sspol_comparison}.\,\textbf{a.} for the $\mathrm{CR_a}$ case(dotted, blue), showing excellent agreement with the numerical solution.

For diffusion rates above approximately $D\sim10^4\,\mathrm{nm\,s^{-1}}$, a scaling of $P_\infty\sim1/D$ is expected for both CR and NOVEL schemes from Eqs.\,\ref{Eq_total_dephasing}\,\&\,\ref{Eq_diff_from_tau_d} due to diffusive motion comprising the dominant contribution to the dephasing. For the case of PulsePol, owing to the discussion in Section\,\ref{Sec_ext_hyperfine}, the reduction in polarisation with increasing $D$ is more severe, exhibiting a scaling of $P_\infty\sim1/D^3$.  

In addition to the realistic decoherence regimes considered above, it is instructive to also consider the potential performance of each scheme under ideal scenarios of negligible NV or target spin decoherence. Repeating the analysis above, but instead using the coherent cooling rates given in Eq.\,\ref{Eq_cool_rate_coherent} reveals that we would expect polarisation yields to improve by an order of magnitude in the case of CR and NOVEL, as shown in Fig.\,\ref{Fig_sspol_comparison}.\,\textbf{b.}. By virtue of its intrinsic dynamical decoupling however, PulsePol would benefit less from this approach, as it is by construction closer to its ideal limit. We have also assumed that the $P_\infty\sim1/D^3\rightarrow 1/D$ scaling could be recovered for pulsed schemes in the ideal case by increasing the nuclear Larmor frequency beyond any diffusion timescale considered here. For example, in the case of water ($D=2.3\times10^{9}\,\mathrm{nm^2s^{-1}}$) we would require $\omega_\mathrm{T}/2\pi\sim800\,\mathrm{MHz}$, which equates to an external field strength of $\sim20\,\mathrm{T}$. The corresponding RF power and precision requirements for pulsing on sub-nanosecond timescales are beyond the reach of most NV research efforts, and would negate the infrastructure cost advantages afforded by the use of a prospective NV-DNP system.
In any case, under the ideal regimes considered for all schemes here, the respective differences in polarisation yield are exclusively due to their respective hyperfine couplings. As such, we expect CR to offer the largest possible yield should fabrication of diamond free of significant surface and bulk electronic spins impurities become achievable.

\section{Conclusion}
In this work we have shown that solid-effect dynamic nuclear polarisation cooling rates can be increased by roughly an order of magnitude via the utilisation of a complementary class of electron-nuclear coupling components from the hyperfine coupling tensor. These components may be readily employed by using an external magnetic field to bring the EPR transition frequencies of spin-$S\geq1$ electrons into resonance with NMR transitions of the target nuclei, thereby achieving direct cross-relaxation (CR) driven nuclear spin cooling\cite{Hal17}. Whilst this method is less robust to electron spin decoherence and requires precise control of the field's magnitude and direction, it avoids many complexities associated with fast pulsing, field inhomogeneities, and power requirements in pulsed-RF methods.

Using the optically spin-polarisable $S=1$ ground state electron spin of the nitrogen-vacancy defect in diamond as an example system, we have experimentally demonstrated this performance increase over the RF-based Hartmann-Hahn (NOVEL) approach\cite{Sha18}, which exhibits the strongest effective hyperfine coupling amongst RF-driven solid-effect DNP methods.
We have developed a theoretical framework to numerically and analytically compare these methods to recently developed DNP schemes based on optimal quantum control, such as PulsePol\cite{Sch18}. In the case where the decoherence of both the electron and nuclei exceed the hyperfine cooling rate, we have demonstrated that despite the in-built electron dynamical decoupling of pulsed schemes, such approaches offer only marginal increases above CR. Moreover, these increases are only exhibited in regimes of low-target motion, such as spin diffusion in solid-state systems. On the other hand, the larger hyperfine couplings associated with the CR approach\cite{Hal17} affords considerable advantage when sources of nuclear spin decoherence, such as nuclear spin dephasing and target motion, exceed that of the electron spin. We note that dynamical decoupling could, in principle, be applied to the target nuclei as well, however implementation on MHz timescales is technically challenging. We have also shown that spin diffusion associated with nuclear dipole couplings amongst protons in organic media offers a near-optimal mixing rate some five to six orders of magnitude below the self diffusion rate of water, and that spatial diffusion should not appreciably exceed this range. This puts solid or gelatinous targets at a considerable advantage over fluidic samples for solid-effect NV-DNP, unless the motion of latter can be somehow constrained.

Future advances in diamond fabrication and surface preparation capable of reducing the densities of electron spin impurities and strain in the diamond would afford CR a significant further advantage over other methods. Optimising the nuclear spin polarisation achievable by individual NV spins may have important applications in polarising macroscopic quantities of nuclei for NMR and MRI, and the exploration of these prospects remains the focus of intense research.

Finally, it is important to note that CR methods do not apply exclusively to the optically polarisable solid-state electron spins studied here, and could be utilised in high-field NMR using transition metal complexes exhibiting zero-field splittings of hundreds of GHz. Whilst beyond the scope of this work, such approaches may find great utility in improving the signal-noise ratio of DNP assisted NMR whilst avoiding the high RF power requirements associated with solid effect DNP.

\newpage
\section{ACKNOWLEDGEMENTS}
We acknowledge support from the Australian Research Council (ARC) through grants FL130100119, DE170100129, CE170100012, and DP190101506.

\bibliography{CRvsHHbib}

\begin{thebibliography}{59}
\expandafter\ifx\csname natexlab\endcsname\relax\def\natexlab#1{#1}\fi
\expandafter\ifx\csname bibnamefont\endcsname\relax
  \def\bibnamefont#1{#1}\fi
\expandafter\ifx\csname bibfnamefont\endcsname\relax
  \def\bibfnamefont#1{#1}\fi
\expandafter\ifx\csname citenamefont\endcsname\relax
  \def\citenamefont#1{#1}\fi
\expandafter\ifx\csname url\endcsname\relax
  \def\url#1{\texttt{#1}}\fi
\expandafter\ifx\csname urlprefix\endcsname\relax\def\urlprefix{URL }\fi
\providecommand{\bibinfo}[2]{#2}
\providecommand{\eprint}[2][]{\url{#2}}

\bibitem[{\citenamefont{Slichter}(1990)}]{Sli}
\bibinfo{author}{\bibfnamefont{C.~P.} \bibnamefont{Slichter}},
  \emph{\bibinfo{title}{Principles of Magnetic Resonance}}
  (\bibinfo{publisher}{Springer}, \bibinfo{address}{Berlin},
  \bibinfo{year}{1990}), \bibinfo{edition}{3rd} ed., \bibinfo{note}{ch.\,7}.

\bibitem[{\citenamefont{Fritsch et~al.}(1991)\citenamefont{Fritsch, Brunner,
  and Hausser}}]{Fri91}
\bibinfo{author}{\bibfnamefont{R.}~\bibnamefont{Fritsch}},
  \bibinfo{author}{\bibfnamefont{H.}~\bibnamefont{Brunner}}, \bibnamefont{and}
  \bibinfo{author}{\bibfnamefont{K.}~\bibnamefont{Hausser}},
  \bibinfo{journal}{Chemical Physics} \textbf{\bibinfo{volume}{151}},
  \bibinfo{pages}{261 } (\bibinfo{year}{1991}).

\bibitem[{\citenamefont{Henstra et~al.}(1990)\citenamefont{Henstra, Lin,
  Schmidt, and Wenckebach}}]{Hen90}
\bibinfo{author}{\bibfnamefont{A.}~\bibnamefont{Henstra}},
  \bibinfo{author}{\bibfnamefont{T.-S.} \bibnamefont{Lin}},
  \bibinfo{author}{\bibfnamefont{J.}~\bibnamefont{Schmidt}}, \bibnamefont{and}
  \bibinfo{author}{\bibfnamefont{W.}~\bibnamefont{Wenckebach}},
  \bibinfo{journal}{Chemical Physics Letters} \textbf{\bibinfo{volume}{165}},
  \bibinfo{pages}{6 } (\bibinfo{year}{1990}).

\bibitem[{\citenamefont{Tateishi et~al.}(2014)\citenamefont{Tateishi, Negoro,
  Nishida, Kagawa, Morita, and Kitagawa}}]{Tat14}
\bibinfo{author}{\bibfnamefont{K.}~\bibnamefont{Tateishi}},
  \bibinfo{author}{\bibfnamefont{M.}~\bibnamefont{Negoro}},
  \bibinfo{author}{\bibfnamefont{S.}~\bibnamefont{Nishida}},
  \bibinfo{author}{\bibfnamefont{A.}~\bibnamefont{Kagawa}},
  \bibinfo{author}{\bibfnamefont{Y.}~\bibnamefont{Morita}}, \bibnamefont{and}
  \bibinfo{author}{\bibfnamefont{M.}~\bibnamefont{Kitagawa}},
  \bibinfo{journal}{Proceedings of the National Academy of Sciences}
  \textbf{\bibinfo{volume}{111}}, \bibinfo{pages}{7527} (\bibinfo{year}{2014}).

\bibitem[{\citenamefont{Doherty et~al.}(2013)\citenamefont{Doherty, Manson,
  Delaney, Jelezko, Wrachtrup, and Hollenberg}}]{Doh13}
\bibinfo{author}{\bibfnamefont{M.~W.} \bibnamefont{Doherty}},
  \bibinfo{author}{\bibfnamefont{N.~B.} \bibnamefont{Manson}},
  \bibinfo{author}{\bibfnamefont{P.}~\bibnamefont{Delaney}},
  \bibinfo{author}{\bibfnamefont{F.}~\bibnamefont{Jelezko}},
  \bibinfo{author}{\bibfnamefont{J.}~\bibnamefont{Wrachtrup}},
  \bibnamefont{and} \bibinfo{author}{\bibfnamefont{L.~C.}
  \bibnamefont{Hollenberg}}, \bibinfo{journal}{Physics Reports}
  \textbf{\bibinfo{volume}{528}}, \bibinfo{pages}{1 } (\bibinfo{year}{2013}).

\bibitem[{\citenamefont{London et~al.}(2013)\citenamefont{London, Scheuer, Cai,
  Schwarz, Retzker, Plenio, Katagiri, Teraji, Koizumi, Isoya et~al.}}]{Lon13}
\bibinfo{author}{\bibfnamefont{P.}~\bibnamefont{London}},
  \bibinfo{author}{\bibfnamefont{J.}~\bibnamefont{Scheuer}},
  \bibinfo{author}{\bibfnamefont{J.~M.} \bibnamefont{Cai}},
  \bibinfo{author}{\bibfnamefont{I.}~\bibnamefont{Schwarz}},
  \bibinfo{author}{\bibfnamefont{A.}~\bibnamefont{Retzker}},
  \bibinfo{author}{\bibfnamefont{M.~B.} \bibnamefont{Plenio}},
  \bibinfo{author}{\bibfnamefont{M.}~\bibnamefont{Katagiri}},
  \bibinfo{author}{\bibfnamefont{T.}~\bibnamefont{Teraji}},
  \bibinfo{author}{\bibfnamefont{S.}~\bibnamefont{Koizumi}},
  \bibinfo{author}{\bibfnamefont{J.}~\bibnamefont{Isoya}},
  \bibnamefont{et~al.}, \bibinfo{journal}{Physical Review Letters}
  \textbf{\bibinfo{volume}{111}}, \bibinfo{pages}{067601}
  (\bibinfo{year}{2013}).

\bibitem[{\citenamefont{Álvarez et~al.}(2015)\citenamefont{Álvarez,
  Bretschneider, Fischer, London, Kanda, Onoda, Isoya, Gershoni, and
  Frydman}}]{Alv15}
\bibinfo{author}{\bibfnamefont{G.~A.} \bibnamefont{Álvarez}},
  \bibinfo{author}{\bibfnamefont{C.~O.} \bibnamefont{Bretschneider}},
  \bibinfo{author}{\bibfnamefont{R.}~\bibnamefont{Fischer}},
  \bibinfo{author}{\bibfnamefont{P.}~\bibnamefont{London}},
  \bibinfo{author}{\bibfnamefont{H.}~\bibnamefont{Kanda}},
  \bibinfo{author}{\bibfnamefont{S.}~\bibnamefont{Onoda}},
  \bibinfo{author}{\bibfnamefont{J.}~\bibnamefont{Isoya}},
  \bibinfo{author}{\bibfnamefont{D.}~\bibnamefont{Gershoni}}, \bibnamefont{and}
  \bibinfo{author}{\bibfnamefont{L.}~\bibnamefont{Frydman}},
  \bibinfo{journal}{Nature Communications} \textbf{\bibinfo{volume}{6}},
  \bibinfo{pages}{8456} (\bibinfo{year}{2015}).

\bibitem[{\citenamefont{King et~al.}(2015)\citenamefont{King, Jeong, Vassiliou,
  Shin, Page, Avalos, Wang, and Pines}}]{kin15}
\bibinfo{author}{\bibfnamefont{J.~P.} \bibnamefont{King}},
  \bibinfo{author}{\bibfnamefont{K.}~\bibnamefont{Jeong}},
  \bibinfo{author}{\bibfnamefont{C.~C.} \bibnamefont{Vassiliou}},
  \bibinfo{author}{\bibfnamefont{C.~S.} \bibnamefont{Shin}},
  \bibinfo{author}{\bibfnamefont{R.~H.} \bibnamefont{Page}},
  \bibinfo{author}{\bibfnamefont{C.~E.} \bibnamefont{Avalos}},
  \bibinfo{author}{\bibfnamefont{H.-J.} \bibnamefont{Wang}}, \bibnamefont{and}
  \bibinfo{author}{\bibfnamefont{A.}~\bibnamefont{Pines}},
  \bibinfo{journal}{Nature Communications} \textbf{\bibinfo{volume}{6}},
  \bibinfo{pages}{8965} (\bibinfo{year}{2015}).

\bibitem[{\citenamefont{Scheuer
  et~al.}(2016{\natexlab{a}})\citenamefont{Scheuer, Schwartz, Chen,
  Schulze-Sünninghausen, Carl, Höfer, Retzker, Sumiya, Isoya, Luy
  et~al.}}]{Sch16}
\bibinfo{author}{\bibfnamefont{J.}~\bibnamefont{Scheuer}},
  \bibinfo{author}{\bibfnamefont{I.}~\bibnamefont{Schwartz}},
  \bibinfo{author}{\bibfnamefont{Q.}~\bibnamefont{Chen}},
  \bibinfo{author}{\bibfnamefont{D.}~\bibnamefont{Schulze-Sünninghausen}},
  \bibinfo{author}{\bibfnamefont{P.}~\bibnamefont{Carl}},
  \bibinfo{author}{\bibfnamefont{P.}~\bibnamefont{Höfer}},
  \bibinfo{author}{\bibfnamefont{A.}~\bibnamefont{Retzker}},
  \bibinfo{author}{\bibfnamefont{H.}~\bibnamefont{Sumiya}},
  \bibinfo{author}{\bibfnamefont{J.}~\bibnamefont{Isoya}},
  \bibinfo{author}{\bibfnamefont{B.}~\bibnamefont{Luy}}, \bibnamefont{et~al.},
  \bibinfo{journal}{New Journal of Physics} \textbf{\bibinfo{volume}{18}},
  \bibinfo{pages}{013040} (\bibinfo{year}{2016}{\natexlab{a}}).

\bibitem[{\citenamefont{Fernández-Acebal
  et~al.}(2018)\citenamefont{Fernández-Acebal, Rosolio, Scheuer, Müller,
  Müller, Schmitt, McGuinness, Schwarz, Chen, Retzker et~al.}}]{Fer18}
\bibinfo{author}{\bibfnamefont{P.}~\bibnamefont{Fernández-Acebal}},
  \bibinfo{author}{\bibfnamefont{O.}~\bibnamefont{Rosolio}},
  \bibinfo{author}{\bibfnamefont{J.}~\bibnamefont{Scheuer}},
  \bibinfo{author}{\bibfnamefont{C.}~\bibnamefont{Müller}},
  \bibinfo{author}{\bibfnamefont{S.}~\bibnamefont{Müller}},
  \bibinfo{author}{\bibfnamefont{S.}~\bibnamefont{Schmitt}},
  \bibinfo{author}{\bibfnamefont{L.}~\bibnamefont{McGuinness}},
  \bibinfo{author}{\bibfnamefont{I.}~\bibnamefont{Schwarz}},
  \bibinfo{author}{\bibfnamefont{Q.}~\bibnamefont{Chen}},
  \bibinfo{author}{\bibfnamefont{A.}~\bibnamefont{Retzker}},
  \bibnamefont{et~al.}, \bibinfo{journal}{Nano Letters}
  \textbf{\bibinfo{volume}{18}}, \bibinfo{pages}{1882} (\bibinfo{year}{2018}).

\bibitem[{\citenamefont{Abrams et~al.}(2014)\citenamefont{Abrams, Trusheim,
  Englund, Shattuck, and Meriles}}]{Abr14}
\bibinfo{author}{\bibfnamefont{D.}~\bibnamefont{Abrams}},
  \bibinfo{author}{\bibfnamefont{M.~E.} \bibnamefont{Trusheim}},
  \bibinfo{author}{\bibfnamefont{D.~R.} \bibnamefont{Englund}},
  \bibinfo{author}{\bibfnamefont{M.~D.} \bibnamefont{Shattuck}},
  \bibnamefont{and} \bibinfo{author}{\bibfnamefont{C.~A.}
  \bibnamefont{Meriles}}, \bibinfo{journal}{Nano Letters}
  \textbf{\bibinfo{volume}{14}}, \bibinfo{pages}{2471} (\bibinfo{year}{2014}).

\bibitem[{\citenamefont{Chen et~al.}(2016)\citenamefont{Chen, Schwarz, Jelezko,
  Retzker, and Plenio}}]{Che16}
\bibinfo{author}{\bibfnamefont{Q.}~\bibnamefont{Chen}},
  \bibinfo{author}{\bibfnamefont{I.}~\bibnamefont{Schwarz}},
  \bibinfo{author}{\bibfnamefont{F.}~\bibnamefont{Jelezko}},
  \bibinfo{author}{\bibfnamefont{A.}~\bibnamefont{Retzker}}, \bibnamefont{and}
  \bibinfo{author}{\bibfnamefont{M.~B.} \bibnamefont{Plenio}},
  \bibinfo{journal}{Physical Review B} \textbf{\bibinfo{volume}{93}},
  \bibinfo{pages}{060408} (\bibinfo{year}{2016}).

\bibitem[{\citenamefont{Scheuer et~al.}(2017)\citenamefont{Scheuer, Schwartz,
  M\"uller, Chen, Dhand, Plenio, Naydenov, and Jelezko}}]{Sch17}
\bibinfo{author}{\bibfnamefont{J.}~\bibnamefont{Scheuer}},
  \bibinfo{author}{\bibfnamefont{I.}~\bibnamefont{Schwartz}},
  \bibinfo{author}{\bibfnamefont{S.}~\bibnamefont{M\"uller}},
  \bibinfo{author}{\bibfnamefont{Q.}~\bibnamefont{Chen}},
  \bibinfo{author}{\bibfnamefont{I.}~\bibnamefont{Dhand}},
  \bibinfo{author}{\bibfnamefont{M.~B.} \bibnamefont{Plenio}},
  \bibinfo{author}{\bibfnamefont{B.}~\bibnamefont{Naydenov}}, \bibnamefont{and}
  \bibinfo{author}{\bibfnamefont{F.}~\bibnamefont{Jelezko}},
  \bibinfo{journal}{Phys. Rev. B} \textbf{\bibinfo{volume}{96}},
  \bibinfo{pages}{174436} (\bibinfo{year}{2017}).

\bibitem[{\citenamefont{Broadway et~al.}(2018)\citenamefont{Broadway, Tetienne,
  Stacey, Wood, Simpson, Hall, and Hollenberg}}]{Hal17}
\bibinfo{author}{\bibfnamefont{D.~A.} \bibnamefont{Broadway}},
  \bibinfo{author}{\bibfnamefont{J.~P.} \bibnamefont{Tetienne}},
  \bibinfo{author}{\bibfnamefont{A.}~\bibnamefont{Stacey}},
  \bibinfo{author}{\bibfnamefont{J.~D.~A.} \bibnamefont{Wood}},
  \bibinfo{author}{\bibfnamefont{D.~A.} \bibnamefont{Simpson}},
  \bibinfo{author}{\bibfnamefont{L.~T.} \bibnamefont{Hall}}, \bibnamefont{and}
  \bibinfo{author}{\bibfnamefont{L.~C.~L.} \bibnamefont{Hollenberg}},
  \bibinfo{journal}{Nature Communications} \textbf{\bibinfo{volume}{9}},
  \bibinfo{pages}{8} (\bibinfo{year}{2018}).

\bibitem[{\citenamefont{Tetienne et~al.}(2020)\citenamefont{Tetienne, Hall,
  Healey, White, A., F., and Hollenberg}}]{Tet20}
\bibinfo{author}{\bibfnamefont{J.~P.} \bibnamefont{Tetienne}},
  \bibinfo{author}{\bibfnamefont{L.~T.} \bibnamefont{Hall}},
  \bibinfo{author}{\bibfnamefont{A.~J.} \bibnamefont{Healey}},
  \bibinfo{author}{\bibfnamefont{G.~A.~L.} \bibnamefont{White}},
  \bibinfo{author}{\bibfnamefont{S.~M.} \bibnamefont{A.}},
  \bibinfo{author}{\bibfnamefont{S.}~\bibnamefont{F.}}, \bibnamefont{and}
  \bibinfo{author}{\bibfnamefont{L.~C.~L.} \bibnamefont{Hollenberg}},
  \bibinfo{journal}{arXiv:2008.12417}  (\bibinfo{year}{2020}).

\bibitem[{\citenamefont{Shagieva et~al.}(2018)\citenamefont{Shagieva, Zaiser,
  Neumann, Dasari, Stöhr, Denisenko, Reuter, Meriles, and Wrachtrup}}]{Sha18}
\bibinfo{author}{\bibfnamefont{F.}~\bibnamefont{Shagieva}},
  \bibinfo{author}{\bibfnamefont{S.}~\bibnamefont{Zaiser}},
  \bibinfo{author}{\bibfnamefont{P.}~\bibnamefont{Neumann}},
  \bibinfo{author}{\bibfnamefont{D.~B.~R.} \bibnamefont{Dasari}},
  \bibinfo{author}{\bibfnamefont{R.}~\bibnamefont{Stöhr}},
  \bibinfo{author}{\bibfnamefont{A.}~\bibnamefont{Denisenko}},
  \bibinfo{author}{\bibfnamefont{R.}~\bibnamefont{Reuter}},
  \bibinfo{author}{\bibfnamefont{C.~A.} \bibnamefont{Meriles}},
  \bibnamefont{and}
  \bibinfo{author}{\bibfnamefont{J.}~\bibnamefont{Wrachtrup}},
  \bibinfo{journal}{Nano Letters} \textbf{\bibinfo{volume}{18}},
  \bibinfo{pages}{3731} (\bibinfo{year}{2018}).

\bibitem[{\citenamefont{Rondin et~al.}(2014)\citenamefont{Rondin, Tetienne,
  Hingant, Roch, Maletinsky, and Jacques}}]{Ron14}
\bibinfo{author}{\bibfnamefont{L.}~\bibnamefont{Rondin}},
  \bibinfo{author}{\bibfnamefont{J.-P.} \bibnamefont{Tetienne}},
  \bibinfo{author}{\bibfnamefont{T.}~\bibnamefont{Hingant}},
  \bibinfo{author}{\bibfnamefont{J.-F.} \bibnamefont{Roch}},
  \bibinfo{author}{\bibfnamefont{P.}~\bibnamefont{Maletinsky}},
  \bibnamefont{and} \bibinfo{author}{\bibfnamefont{V.}~\bibnamefont{Jacques}},
  \bibinfo{journal}{Reports on Progress in Physics}
  \textbf{\bibinfo{volume}{77}}, \bibinfo{pages}{056503}
  (\bibinfo{year}{2014}).

\bibitem[{\citenamefont{Schirhagl et~al.}(2014)\citenamefont{Schirhagl, Chang,
  Loretz, and Degen}}]{Sch14}
\bibinfo{author}{\bibfnamefont{R.}~\bibnamefont{Schirhagl}},
  \bibinfo{author}{\bibfnamefont{K.}~\bibnamefont{Chang}},
  \bibinfo{author}{\bibfnamefont{M.}~\bibnamefont{Loretz}}, \bibnamefont{and}
  \bibinfo{author}{\bibfnamefont{C.~L.} \bibnamefont{Degen}},
  \bibinfo{journal}{Annual Review of Physical Chemistry}
  \textbf{\bibinfo{volume}{65}}, \bibinfo{pages}{83} (\bibinfo{year}{2014}).

\bibitem[{\citenamefont{Casola et~al.}(2018)\citenamefont{Casola, van~der Sar,
  and Yacoby}}]{Cas18}
\bibinfo{author}{\bibfnamefont{F.}~\bibnamefont{Casola}},
  \bibinfo{author}{\bibfnamefont{T.}~\bibnamefont{van~der Sar}},
  \bibnamefont{and} \bibinfo{author}{\bibfnamefont{A.}~\bibnamefont{Yacoby}},
  \bibinfo{journal}{Nature Reviews Materials} \textbf{\bibinfo{volume}{3}},
  \bibinfo{pages}{17088} (\bibinfo{year}{2018}).

\bibitem[{\citenamefont{Waldherr et~al.}(2011)\citenamefont{Waldherr, Beck,
  Steiner, Neumann, Gali, Frauenheim, Jelezko, and Wrachtrup}}]{Wal11}
\bibinfo{author}{\bibfnamefont{G.}~\bibnamefont{Waldherr}},
  \bibinfo{author}{\bibfnamefont{J.}~\bibnamefont{Beck}},
  \bibinfo{author}{\bibfnamefont{M.}~\bibnamefont{Steiner}},
  \bibinfo{author}{\bibfnamefont{P.}~\bibnamefont{Neumann}},
  \bibinfo{author}{\bibfnamefont{A.}~\bibnamefont{Gali}},
  \bibinfo{author}{\bibfnamefont{T.}~\bibnamefont{Frauenheim}},
  \bibinfo{author}{\bibfnamefont{F.}~\bibnamefont{Jelezko}}, \bibnamefont{and}
  \bibinfo{author}{\bibfnamefont{J.}~\bibnamefont{Wrachtrup}},
  \bibinfo{journal}{Physical Review Letters} \textbf{\bibinfo{volume}{106}},
  \bibinfo{pages}{157601} (\bibinfo{year}{2011}).

\bibitem[{\citenamefont{Overhauser}(1953)}]{Ove53}
\bibinfo{author}{\bibfnamefont{A.~W.} \bibnamefont{Overhauser}},
  \bibinfo{journal}{Physical Review} \textbf{\bibinfo{volume}{92}},
  \bibinfo{pages}{411} (\bibinfo{year}{1953}).

\bibitem[{\citenamefont{Jeffries}(1957)}]{Jef57}
\bibinfo{author}{\bibfnamefont{C.~D.} \bibnamefont{Jeffries}},
  \bibinfo{journal}{Physical Review} \textbf{\bibinfo{volume}{106}},
  \bibinfo{pages}{164} (\bibinfo{year}{1957}).

\bibitem[{\citenamefont{Hwang and Hill}(1967)}]{Hwa67}
\bibinfo{author}{\bibfnamefont{C.~F.} \bibnamefont{Hwang}} \bibnamefont{and}
  \bibinfo{author}{\bibfnamefont{D.~A.} \bibnamefont{Hill}},
  \bibinfo{journal}{Phys. Rev. Lett.} \textbf{\bibinfo{volume}{18}},
  \bibinfo{pages}{110} (\bibinfo{year}{1967}).

\bibitem[{\citenamefont{Abragam and Goldman}(1978)}]{Abr78}
\bibinfo{author}{\bibfnamefont{A.}~\bibnamefont{Abragam}} \bibnamefont{and}
  \bibinfo{author}{\bibfnamefont{M.}~\bibnamefont{Goldman}},
  \bibinfo{journal}{Reports on Progress in Physics}
  \textbf{\bibinfo{volume}{41}}, \bibinfo{pages}{395} (\bibinfo{year}{1978}).

\bibitem[{\citenamefont{King et~al.}(2010)\citenamefont{King, Coles, and
  Reimer}}]{Kin10}
\bibinfo{author}{\bibfnamefont{J.~P.} \bibnamefont{King}},
  \bibinfo{author}{\bibfnamefont{P.~J.} \bibnamefont{Coles}}, \bibnamefont{and}
  \bibinfo{author}{\bibfnamefont{J.~A.} \bibnamefont{Reimer}},
  \bibinfo{journal}{Physical Review B} \textbf{\bibinfo{volume}{81}},
  \bibinfo{pages}{073201} (\bibinfo{year}{2010}).

\bibitem[{\citenamefont{Pagliero et~al.}(2018)\citenamefont{Pagliero, Rao,
  Zangara, Dhomkar, Wong, Abril, Aslam, Parker, King, Avalos et~al.}}]{Pag18}
\bibinfo{author}{\bibfnamefont{D.}~\bibnamefont{Pagliero}},
  \bibinfo{author}{\bibfnamefont{K.~R.~K.} \bibnamefont{Rao}},
  \bibinfo{author}{\bibfnamefont{P.~R.} \bibnamefont{Zangara}},
  \bibinfo{author}{\bibfnamefont{S.}~\bibnamefont{Dhomkar}},
  \bibinfo{author}{\bibfnamefont{H.~H.} \bibnamefont{Wong}},
  \bibinfo{author}{\bibfnamefont{A.}~\bibnamefont{Abril}},
  \bibinfo{author}{\bibfnamefont{N.}~\bibnamefont{Aslam}},
  \bibinfo{author}{\bibfnamefont{A.}~\bibnamefont{Parker}},
  \bibinfo{author}{\bibfnamefont{J.}~\bibnamefont{King}},
  \bibinfo{author}{\bibfnamefont{C.~E.} \bibnamefont{Avalos}},
  \bibnamefont{et~al.}, \bibinfo{journal}{Phys. Rev. B}
  \textbf{\bibinfo{volume}{97}}, \bibinfo{pages}{024422}
  (\bibinfo{year}{2018}).

\bibitem[{\citenamefont{Henshaw et~al.}(2019)\citenamefont{Henshaw, Pagliero,
  Zangara, Franzoni, Ajoy, Acosta, Reimer, Pines, and Meriles}}]{Hen19}
\bibinfo{author}{\bibfnamefont{J.}~\bibnamefont{Henshaw}},
  \bibinfo{author}{\bibfnamefont{D.}~\bibnamefont{Pagliero}},
  \bibinfo{author}{\bibfnamefont{P.~R.} \bibnamefont{Zangara}},
  \bibinfo{author}{\bibfnamefont{M.~B.} \bibnamefont{Franzoni}},
  \bibinfo{author}{\bibfnamefont{A.}~\bibnamefont{Ajoy}},
  \bibinfo{author}{\bibfnamefont{R.~H.} \bibnamefont{Acosta}},
  \bibinfo{author}{\bibfnamefont{J.~A.} \bibnamefont{Reimer}},
  \bibinfo{author}{\bibfnamefont{A.}~\bibnamefont{Pines}}, \bibnamefont{and}
  \bibinfo{author}{\bibfnamefont{C.~A.} \bibnamefont{Meriles}},
  \bibinfo{journal}{Proceedings of the National Academy of Sciences}
  \textbf{\bibinfo{volume}{116}}, \bibinfo{pages}{18334}
  (\bibinfo{year}{2019}).

\bibitem[{\citenamefont{Jacques et~al.}(2009)\citenamefont{Jacques, Neumann,
  Beck, Markham, Twitchen, Meijer, Kaiser, Balasubramanian, Jelezko, and
  Wrachtrup}}]{Jac09}
\bibinfo{author}{\bibfnamefont{V.}~\bibnamefont{Jacques}},
  \bibinfo{author}{\bibfnamefont{P.}~\bibnamefont{Neumann}},
  \bibinfo{author}{\bibfnamefont{J.}~\bibnamefont{Beck}},
  \bibinfo{author}{\bibfnamefont{M.}~\bibnamefont{Markham}},
  \bibinfo{author}{\bibfnamefont{D.}~\bibnamefont{Twitchen}},
  \bibinfo{author}{\bibfnamefont{J.}~\bibnamefont{Meijer}},
  \bibinfo{author}{\bibfnamefont{F.}~\bibnamefont{Kaiser}},
  \bibinfo{author}{\bibfnamefont{G.}~\bibnamefont{Balasubramanian}},
  \bibinfo{author}{\bibfnamefont{F.}~\bibnamefont{Jelezko}}, \bibnamefont{and}
  \bibinfo{author}{\bibfnamefont{J.}~\bibnamefont{Wrachtrup}},
  \bibinfo{journal}{Physical Review Letters} \textbf{\bibinfo{volume}{102}},
  \bibinfo{pages}{057403} (\bibinfo{year}{2009}).

\bibitem[{\citenamefont{Gruber et~al.}(1997)\citenamefont{Gruber, Dräbenstedt,
  Tietz, Fleury, Wrachtrup, and Borczyskowski}}]{Gru97}
\bibinfo{author}{\bibfnamefont{A.}~\bibnamefont{Gruber}},
  \bibinfo{author}{\bibfnamefont{A.}~\bibnamefont{Dräbenstedt}},
  \bibinfo{author}{\bibfnamefont{C.}~\bibnamefont{Tietz}},
  \bibinfo{author}{\bibfnamefont{L.}~\bibnamefont{Fleury}},
  \bibinfo{author}{\bibfnamefont{J.}~\bibnamefont{Wrachtrup}},
  \bibnamefont{and} \bibinfo{author}{\bibfnamefont{C.~v.}
  \bibnamefont{Borczyskowski}}, \bibinfo{journal}{Science}
  \textbf{\bibinfo{volume}{276}}, \bibinfo{pages}{2012} (\bibinfo{year}{1997}).

\bibitem[{\citenamefont{Fischer et~al.}(2013)\citenamefont{Fischer,
  Bretschneider, London, Budker, Gershoni, and Frydman}}]{Fis13}
\bibinfo{author}{\bibfnamefont{R.}~\bibnamefont{Fischer}},
  \bibinfo{author}{\bibfnamefont{C.~O.} \bibnamefont{Bretschneider}},
  \bibinfo{author}{\bibfnamefont{P.}~\bibnamefont{London}},
  \bibinfo{author}{\bibfnamefont{D.}~\bibnamefont{Budker}},
  \bibinfo{author}{\bibfnamefont{D.}~\bibnamefont{Gershoni}}, \bibnamefont{and}
  \bibinfo{author}{\bibfnamefont{L.}~\bibnamefont{Frydman}},
  \bibinfo{journal}{Physical Review Letters} \textbf{\bibinfo{volume}{111}},
  \bibinfo{pages}{057601} (\bibinfo{year}{2013}).

\bibitem[{\citenamefont{Wunderlich et~al.}(2017)\citenamefont{Wunderlich,
  Kohlrautz, Abel, Haase, and Meijer}}]{Wun17}
\bibinfo{author}{\bibfnamefont{R.}~\bibnamefont{Wunderlich}},
  \bibinfo{author}{\bibfnamefont{J.}~\bibnamefont{Kohlrautz}},
  \bibinfo{author}{\bibfnamefont{B.}~\bibnamefont{Abel}},
  \bibinfo{author}{\bibfnamefont{J.}~\bibnamefont{Haase}}, \bibnamefont{and}
  \bibinfo{author}{\bibfnamefont{J.}~\bibnamefont{Meijer}},
  \bibinfo{journal}{Physical Review B} \textbf{\bibinfo{volume}{96}},
  \bibinfo{pages}{220407} (\bibinfo{year}{2017}).

\bibitem[{\citenamefont{Wang et~al.}(2013)\citenamefont{Wang, Shin, Avalos,
  Seltzer, Budker, Pines, and Bajaj}}]{Wan13}
\bibinfo{author}{\bibfnamefont{H.-J.} \bibnamefont{Wang}},
  \bibinfo{author}{\bibfnamefont{C.~S.} \bibnamefont{Shin}},
  \bibinfo{author}{\bibfnamefont{C.~E.} \bibnamefont{Avalos}},
  \bibinfo{author}{\bibfnamefont{S.~J.} \bibnamefont{Seltzer}},
  \bibinfo{author}{\bibfnamefont{D.}~\bibnamefont{Budker}},
  \bibinfo{author}{\bibfnamefont{A.}~\bibnamefont{Pines}}, \bibnamefont{and}
  \bibinfo{author}{\bibfnamefont{V.~S.} \bibnamefont{Bajaj}},
  \bibinfo{journal}{Nature Communications} \textbf{\bibinfo{volume}{4}},
  \bibinfo{pages}{1940} (\bibinfo{year}{2013}).

\bibitem[{\citenamefont{Chen et~al.}(2015)\citenamefont{Chen, Schwarz, Jelezko,
  Retzker, and Plenio}}]{Che15}
\bibinfo{author}{\bibfnamefont{Q.}~\bibnamefont{Chen}},
  \bibinfo{author}{\bibfnamefont{I.}~\bibnamefont{Schwarz}},
  \bibinfo{author}{\bibfnamefont{F.}~\bibnamefont{Jelezko}},
  \bibinfo{author}{\bibfnamefont{A.}~\bibnamefont{Retzker}}, \bibnamefont{and}
  \bibinfo{author}{\bibfnamefont{M.~B.} \bibnamefont{Plenio}},
  \bibinfo{journal}{Physical Review B} \textbf{\bibinfo{volume}{92}},
  \bibinfo{pages}{184420} (\bibinfo{year}{2015}).

\bibitem[{\citenamefont{Scheuer
  et~al.}(2016{\natexlab{b}})\citenamefont{Scheuer, Schwartz, Chen,
  Schulze-Sünninghausen, Carl, Höfer, Retzker, Sumiya, Isoya, Luy
  et~al.}}]{Joc16}
\bibinfo{author}{\bibfnamefont{J.}~\bibnamefont{Scheuer}},
  \bibinfo{author}{\bibfnamefont{I.}~\bibnamefont{Schwartz}},
  \bibinfo{author}{\bibfnamefont{Q.}~\bibnamefont{Chen}},
  \bibinfo{author}{\bibfnamefont{D.}~\bibnamefont{Schulze-Sünninghausen}},
  \bibinfo{author}{\bibfnamefont{P.}~\bibnamefont{Carl}},
  \bibinfo{author}{\bibfnamefont{P.}~\bibnamefont{Höfer}},
  \bibinfo{author}{\bibfnamefont{A.}~\bibnamefont{Retzker}},
  \bibinfo{author}{\bibfnamefont{H.}~\bibnamefont{Sumiya}},
  \bibinfo{author}{\bibfnamefont{J.}~\bibnamefont{Isoya}},
  \bibinfo{author}{\bibfnamefont{B.}~\bibnamefont{Luy}}, \bibnamefont{et~al.},
  \bibinfo{journal}{New Journal of Physics} \textbf{\bibinfo{volume}{18}},
  \bibinfo{pages}{013040} (\bibinfo{year}{2016}{\natexlab{b}}).

\bibitem[{\citenamefont{Ajoy et~al.}(2018)\citenamefont{Ajoy, Liu, Nazaryan,
  Lv, Zangara, Safvati, Wang, Arnold, Li, Lin et~al.}}]{Ajo18}
\bibinfo{author}{\bibfnamefont{A.}~\bibnamefont{Ajoy}},
  \bibinfo{author}{\bibfnamefont{K.}~\bibnamefont{Liu}},
  \bibinfo{author}{\bibfnamefont{R.}~\bibnamefont{Nazaryan}},
  \bibinfo{author}{\bibfnamefont{X.}~\bibnamefont{Lv}},
  \bibinfo{author}{\bibfnamefont{P.~R.} \bibnamefont{Zangara}},
  \bibinfo{author}{\bibfnamefont{B.}~\bibnamefont{Safvati}},
  \bibinfo{author}{\bibfnamefont{G.}~\bibnamefont{Wang}},
  \bibinfo{author}{\bibfnamefont{D.}~\bibnamefont{Arnold}},
  \bibinfo{author}{\bibfnamefont{G.}~\bibnamefont{Li}},
  \bibinfo{author}{\bibfnamefont{A.}~\bibnamefont{Lin}}, \bibnamefont{et~al.},
  \bibinfo{journal}{Science Advances} \textbf{\bibinfo{volume}{4}}
  (\bibinfo{year}{2018}).

\bibitem[{\citenamefont{Lang et~al.}(2019{\natexlab{a}})\citenamefont{Lang,
  Broadway, White, Hall, Stacey, Hollenberg, Monteiro, and Tetienne}}]{Lan19}
\bibinfo{author}{\bibfnamefont{J.~E.} \bibnamefont{Lang}},
  \bibinfo{author}{\bibfnamefont{D.~A.} \bibnamefont{Broadway}},
  \bibinfo{author}{\bibfnamefont{G.~A.~L.} \bibnamefont{White}},
  \bibinfo{author}{\bibfnamefont{L.~T.} \bibnamefont{Hall}},
  \bibinfo{author}{\bibfnamefont{A.}~\bibnamefont{Stacey}},
  \bibinfo{author}{\bibfnamefont{L.~C.~L.} \bibnamefont{Hollenberg}},
  \bibinfo{author}{\bibfnamefont{T.~S.} \bibnamefont{Monteiro}},
  \bibnamefont{and} \bibinfo{author}{\bibfnamefont{J.-P.}
  \bibnamefont{Tetienne}}, \bibinfo{journal}{Phys. Rev. Lett.}
  \textbf{\bibinfo{volume}{123}}, \bibinfo{pages}{210401}
  (\bibinfo{year}{2019}{\natexlab{a}}).

\bibitem[{\citenamefont{Tan et~al.}(2019)\citenamefont{Tan, Yang, Weber,
  Mathies, and Griffin}}]{Tan20}
\bibinfo{author}{\bibfnamefont{K.~O.} \bibnamefont{Tan}},
  \bibinfo{author}{\bibfnamefont{C.}~\bibnamefont{Yang}},
  \bibinfo{author}{\bibfnamefont{R.~T.} \bibnamefont{Weber}},
  \bibinfo{author}{\bibfnamefont{G.}~\bibnamefont{Mathies}}, \bibnamefont{and}
  \bibinfo{author}{\bibfnamefont{R.~G.} \bibnamefont{Griffin}},
  \bibinfo{journal}{Science Advances} \textbf{\bibinfo{volume}{5}}
  (\bibinfo{year}{2019}).

\bibitem[{\citenamefont{Schwartz et~al.}(2018)\citenamefont{Schwartz, Scheuer,
  Tratzmiller, M{\"u}ller, Chen, Dhand, Wang, M{\"u}ller, Naydenov, Jelezko
  et~al.}}]{Sch18}
\bibinfo{author}{\bibfnamefont{I.}~\bibnamefont{Schwartz}},
  \bibinfo{author}{\bibfnamefont{J.}~\bibnamefont{Scheuer}},
  \bibinfo{author}{\bibfnamefont{B.}~\bibnamefont{Tratzmiller}},
  \bibinfo{author}{\bibfnamefont{S.}~\bibnamefont{M{\"u}ller}},
  \bibinfo{author}{\bibfnamefont{Q.}~\bibnamefont{Chen}},
  \bibinfo{author}{\bibfnamefont{I.}~\bibnamefont{Dhand}},
  \bibinfo{author}{\bibfnamefont{Z.-Y.} \bibnamefont{Wang}},
  \bibinfo{author}{\bibfnamefont{C.}~\bibnamefont{M{\"u}ller}},
  \bibinfo{author}{\bibfnamefont{B.}~\bibnamefont{Naydenov}},
  \bibinfo{author}{\bibfnamefont{F.}~\bibnamefont{Jelezko}},
  \bibnamefont{et~al.}, \bibinfo{journal}{Science Advances}
  \textbf{\bibinfo{volume}{4}} (\bibinfo{year}{2018}).

\bibitem[{\citenamefont{Broadway et~al.}(2016)\citenamefont{Broadway, Wood,
  Hall, Stacey, Markham, Simpson, Tetienne, and Hollenberg}}]{Bro16}
\bibinfo{author}{\bibfnamefont{D.~A.} \bibnamefont{Broadway}},
  \bibinfo{author}{\bibfnamefont{J.~D.~A.} \bibnamefont{Wood}},
  \bibinfo{author}{\bibfnamefont{L.~T.} \bibnamefont{Hall}},
  \bibinfo{author}{\bibfnamefont{A.}~\bibnamefont{Stacey}},
  \bibinfo{author}{\bibfnamefont{M.}~\bibnamefont{Markham}},
  \bibinfo{author}{\bibfnamefont{D.~A.} \bibnamefont{Simpson}},
  \bibinfo{author}{\bibfnamefont{J.~P.} \bibnamefont{Tetienne}},
  \bibnamefont{and} \bibinfo{author}{\bibfnamefont{L.~C.~L.}
  \bibnamefont{Hollenberg}}, \bibinfo{journal}{Physical Review Applied}
  \textbf{\bibinfo{volume}{6}}, \bibinfo{pages}{11} (\bibinfo{year}{2016}).

\bibitem[{\citenamefont{Hall et~al.}(2016)\citenamefont{Hall, Kehayias,
  Simpson, Jarmola, Stacey, Budker, and Hollenberg}}]{Hal16}
\bibinfo{author}{\bibfnamefont{L.~T.} \bibnamefont{Hall}},
  \bibinfo{author}{\bibfnamefont{P.}~\bibnamefont{Kehayias}},
  \bibinfo{author}{\bibfnamefont{D.~A.} \bibnamefont{Simpson}},
  \bibinfo{author}{\bibfnamefont{A.}~\bibnamefont{Jarmola}},
  \bibinfo{author}{\bibfnamefont{A.}~\bibnamefont{Stacey}},
  \bibinfo{author}{\bibfnamefont{D.}~\bibnamefont{Budker}}, \bibnamefont{and}
  \bibinfo{author}{\bibfnamefont{L.~C.~L.} \bibnamefont{Hollenberg}},
  \bibinfo{journal}{Nature communications} \textbf{\bibinfo{volume}{7}},
  \bibinfo{pages}{10211} (\bibinfo{year}{2016}).

\bibitem[{\citenamefont{Wood et~al.}(2017)\citenamefont{Wood, Tetienne,
  Broadway, Hall, Simpson, Stacey, and Hollenberg}}]{Woo17}
\bibinfo{author}{\bibfnamefont{J.~D.~A.} \bibnamefont{Wood}},
  \bibinfo{author}{\bibfnamefont{J.-P.} \bibnamefont{Tetienne}},
  \bibinfo{author}{\bibfnamefont{D.~A.} \bibnamefont{Broadway}},
  \bibinfo{author}{\bibfnamefont{L.~T.} \bibnamefont{Hall}},
  \bibinfo{author}{\bibfnamefont{D.~A.} \bibnamefont{Simpson}},
  \bibinfo{author}{\bibfnamefont{A.}~\bibnamefont{Stacey}}, \bibnamefont{and}
  \bibinfo{author}{\bibfnamefont{L.~C.~L.} \bibnamefont{Hollenberg}},
  \bibinfo{journal}{Nature Communications} \textbf{\bibinfo{volume}{8}},
  \bibinfo{pages}{15950} (\bibinfo{year}{2017}).

\bibitem[{\citenamefont{Cywinski et~al.}(2009)\citenamefont{Cywinski, Witzel,
  and Das~Sarma}}]{Cyw09}
\bibinfo{author}{\bibfnamefont{L.}~\bibnamefont{Cywinski}},
  \bibinfo{author}{\bibfnamefont{W.~M.} \bibnamefont{Witzel}},
  \bibnamefont{and}
  \bibinfo{author}{\bibfnamefont{S.}~\bibnamefont{Das~Sarma}},
  \bibinfo{journal}{Physical Review Letters} \textbf{\bibinfo{volume}{102}}
  (\bibinfo{year}{2009}).

\bibitem[{\citenamefont{Mamin et~al.}(2013)\citenamefont{Mamin, Kim, Sherwood,
  Rettner, Ohno, Awschalom, and Rugar}}]{Mam13}
\bibinfo{author}{\bibfnamefont{H.~J.} \bibnamefont{Mamin}},
  \bibinfo{author}{\bibfnamefont{M.}~\bibnamefont{Kim}},
  \bibinfo{author}{\bibfnamefont{M.~H.} \bibnamefont{Sherwood}},
  \bibinfo{author}{\bibfnamefont{C.~T.} \bibnamefont{Rettner}},
  \bibinfo{author}{\bibfnamefont{K.}~\bibnamefont{Ohno}},
  \bibinfo{author}{\bibfnamefont{D.~D.} \bibnamefont{Awschalom}},
  \bibnamefont{and} \bibinfo{author}{\bibfnamefont{D.}~\bibnamefont{Rugar}},
  \bibinfo{journal}{Science} \textbf{\bibinfo{volume}{339}},
  \bibinfo{pages}{557} (\bibinfo{year}{2013}).

\bibitem[{\citenamefont{Staudacher et~al.}(2013)\citenamefont{Staudacher, Shi,
  Pezzagna, Meijer, Du, Meriles, Reinhard, and Wrachtrup}}]{Stu13}
\bibinfo{author}{\bibfnamefont{T.}~\bibnamefont{Staudacher}},
  \bibinfo{author}{\bibfnamefont{F.}~\bibnamefont{Shi}},
  \bibinfo{author}{\bibfnamefont{S.}~\bibnamefont{Pezzagna}},
  \bibinfo{author}{\bibfnamefont{J.}~\bibnamefont{Meijer}},
  \bibinfo{author}{\bibfnamefont{J.}~\bibnamefont{Du}},
  \bibinfo{author}{\bibfnamefont{C.~A.} \bibnamefont{Meriles}},
  \bibinfo{author}{\bibfnamefont{F.}~\bibnamefont{Reinhard}}, \bibnamefont{and}
  \bibinfo{author}{\bibfnamefont{J.}~\bibnamefont{Wrachtrup}},
  \bibinfo{journal}{Science} \textbf{\bibinfo{volume}{339}},
  \bibinfo{pages}{561} (\bibinfo{year}{2013}).

\bibitem[{\citenamefont{Hatada et~al.}(1976)\citenamefont{Hatada, Okamoto,
  Ohta, and Yuki}}]{Hat76}
\bibinfo{author}{\bibfnamefont{K.}~\bibnamefont{Hatada}},
  \bibinfo{author}{\bibfnamefont{Y.}~\bibnamefont{Okamoto}},
  \bibinfo{author}{\bibfnamefont{K.}~\bibnamefont{Ohta}}, \bibnamefont{and}
  \bibinfo{author}{\bibfnamefont{H.}~\bibnamefont{Yuki}},
  \bibinfo{journal}{Journal of Polymer Science: Polymer Letters Edition}
  \textbf{\bibinfo{volume}{14}}, \bibinfo{pages}{51} (\bibinfo{year}{1976}).

\bibitem[{\citenamefont{Sangtawesin et~al.}(2019)\citenamefont{Sangtawesin,
  Dwyer, Srinivasan, Allred, Rodgers, De~Greve, Stacey, Dontschuk, O'Donnell,
  Hu et~al.}}]{San19}
\bibinfo{author}{\bibfnamefont{S.}~\bibnamefont{Sangtawesin}},
  \bibinfo{author}{\bibfnamefont{B.~L.} \bibnamefont{Dwyer}},
  \bibinfo{author}{\bibfnamefont{S.}~\bibnamefont{Srinivasan}},
  \bibinfo{author}{\bibfnamefont{J.~J.} \bibnamefont{Allred}},
  \bibinfo{author}{\bibfnamefont{L.~V.~H.} \bibnamefont{Rodgers}},
  \bibinfo{author}{\bibfnamefont{K.}~\bibnamefont{De~Greve}},
  \bibinfo{author}{\bibfnamefont{A.}~\bibnamefont{Stacey}},
  \bibinfo{author}{\bibfnamefont{N.}~\bibnamefont{Dontschuk}},
  \bibinfo{author}{\bibfnamefont{K.~M.} \bibnamefont{O'Donnell}},
  \bibinfo{author}{\bibfnamefont{D.}~\bibnamefont{Hu}}, \bibnamefont{et~al.},
  \bibinfo{journal}{Phys. Rev. X} \textbf{\bibinfo{volume}{9}},
  \bibinfo{pages}{031052} (\bibinfo{year}{2019}).

\bibitem[{\citenamefont{Hanson et~al.}(2008)\citenamefont{Hanson, Dobrovitski,
  Feiguin, Gywat, and Awschalom}}]{Han08}
\bibinfo{author}{\bibfnamefont{R.}~\bibnamefont{Hanson}},
  \bibinfo{author}{\bibfnamefont{V.~V.} \bibnamefont{Dobrovitski}},
  \bibinfo{author}{\bibfnamefont{A.~E.} \bibnamefont{Feiguin}},
  \bibinfo{author}{\bibfnamefont{O.}~\bibnamefont{Gywat}}, \bibnamefont{and}
  \bibinfo{author}{\bibfnamefont{D.~D.} \bibnamefont{Awschalom}},
  \bibinfo{journal}{Science} \textbf{\bibinfo{volume}{320}},
  \bibinfo{pages}{352} (\bibinfo{year}{2008}).

\bibitem[{\citenamefont{Grinolds et~al.}(2014)\citenamefont{Grinolds, Warner,
  De~Greve, Dovzhenko, Thiel, Walsworth, Hong, Maletinsky, and Yacoby}}]{Gri14}
\bibinfo{author}{\bibfnamefont{M.~S.} \bibnamefont{Grinolds}},
  \bibinfo{author}{\bibfnamefont{M.}~\bibnamefont{Warner}},
  \bibinfo{author}{\bibfnamefont{K.}~\bibnamefont{De~Greve}},
  \bibinfo{author}{\bibfnamefont{Y.}~\bibnamefont{Dovzhenko}},
  \bibinfo{author}{\bibfnamefont{L.}~\bibnamefont{Thiel}},
  \bibinfo{author}{\bibfnamefont{R.~L.} \bibnamefont{Walsworth}},
  \bibinfo{author}{\bibfnamefont{S.}~\bibnamefont{Hong}},
  \bibinfo{author}{\bibfnamefont{P.}~\bibnamefont{Maletinsky}},
  \bibnamefont{and} \bibinfo{author}{\bibfnamefont{A.}~\bibnamefont{Yacoby}},
  \bibinfo{journal}{Nature Nanotechnology} \textbf{\bibinfo{volume}{9}},
  \bibinfo{pages}{279} (\bibinfo{year}{2014}).

\bibitem[{\citenamefont{Rosskopf et~al.}(2014)\citenamefont{Rosskopf, Dussaux,
  Ohashi, Loretz, Schirhagl, Watanabe, Shikata, Itoh, and Degen}}]{Ros14}
\bibinfo{author}{\bibfnamefont{T.}~\bibnamefont{Rosskopf}},
  \bibinfo{author}{\bibfnamefont{A.}~\bibnamefont{Dussaux}},
  \bibinfo{author}{\bibfnamefont{K.}~\bibnamefont{Ohashi}},
  \bibinfo{author}{\bibfnamefont{M.}~\bibnamefont{Loretz}},
  \bibinfo{author}{\bibfnamefont{R.}~\bibnamefont{Schirhagl}},
  \bibinfo{author}{\bibfnamefont{H.}~\bibnamefont{Watanabe}},
  \bibinfo{author}{\bibfnamefont{S.}~\bibnamefont{Shikata}},
  \bibinfo{author}{\bibfnamefont{K.~M.} \bibnamefont{Itoh}}, \bibnamefont{and}
  \bibinfo{author}{\bibfnamefont{C.~L.} \bibnamefont{Degen}},
  \bibinfo{journal}{Phys. Rev. Lett.} \textbf{\bibinfo{volume}{112}},
  \bibinfo{pages}{147602} (\bibinfo{year}{2014}).

\bibitem[{\citenamefont{Myers et~al.}(2014)\citenamefont{Myers, Das, Dartiailh,
  Ohno, Awschalom, and Bleszynski~Jayich}}]{Mye14}
\bibinfo{author}{\bibfnamefont{B.~A.} \bibnamefont{Myers}},
  \bibinfo{author}{\bibfnamefont{A.}~\bibnamefont{Das}},
  \bibinfo{author}{\bibfnamefont{M.~C.} \bibnamefont{Dartiailh}},
  \bibinfo{author}{\bibfnamefont{K.}~\bibnamefont{Ohno}},
  \bibinfo{author}{\bibfnamefont{D.~D.} \bibnamefont{Awschalom}},
  \bibnamefont{and} \bibinfo{author}{\bibfnamefont{A.~C.}
  \bibnamefont{Bleszynski~Jayich}}, \bibinfo{journal}{Phys. Rev. Lett.}
  \textbf{\bibinfo{volume}{113}}, \bibinfo{pages}{027602}
  (\bibinfo{year}{2014}).

\bibitem[{\citenamefont{Luan et~al.}(2015)\citenamefont{Luan, Grinolds, Hong,
  Maletinsky, Walsworth, and Yacoby}}]{Lua15}
\bibinfo{author}{\bibfnamefont{L.}~\bibnamefont{Luan}},
  \bibinfo{author}{\bibfnamefont{M.~S.} \bibnamefont{Grinolds}},
  \bibinfo{author}{\bibfnamefont{S.}~\bibnamefont{Hong}},
  \bibinfo{author}{\bibfnamefont{P.}~\bibnamefont{Maletinsky}},
  \bibinfo{author}{\bibfnamefont{R.~L.} \bibnamefont{Walsworth}},
  \bibnamefont{and} \bibinfo{author}{\bibfnamefont{A.}~\bibnamefont{Yacoby}},
  \bibinfo{journal}{Scientific Reports} \textbf{\bibinfo{volume}{5}},
  \bibinfo{pages}{8119} (\bibinfo{year}{2015}).

\bibitem[{\citenamefont{Romach et~al.}(2015)\citenamefont{Romach, M\"uller,
  Unden, Rogers, Isoda, Itoh, Markham, Stacey, Meijer, Pezzagna
  et~al.}}]{Rom15}
\bibinfo{author}{\bibfnamefont{Y.}~\bibnamefont{Romach}},
  \bibinfo{author}{\bibfnamefont{C.}~\bibnamefont{M\"uller}},
  \bibinfo{author}{\bibfnamefont{T.}~\bibnamefont{Unden}},
  \bibinfo{author}{\bibfnamefont{L.~J.} \bibnamefont{Rogers}},
  \bibinfo{author}{\bibfnamefont{T.}~\bibnamefont{Isoda}},
  \bibinfo{author}{\bibfnamefont{K.~M.} \bibnamefont{Itoh}},
  \bibinfo{author}{\bibfnamefont{M.}~\bibnamefont{Markham}},
  \bibinfo{author}{\bibfnamefont{A.}~\bibnamefont{Stacey}},
  \bibinfo{author}{\bibfnamefont{J.}~\bibnamefont{Meijer}},
  \bibinfo{author}{\bibfnamefont{S.}~\bibnamefont{Pezzagna}},
  \bibnamefont{et~al.}, \bibinfo{journal}{Phys. Rev. Lett.}
  \textbf{\bibinfo{volume}{114}}, \bibinfo{pages}{017601}
  (\bibinfo{year}{2015}).

\bibitem[{\citenamefont{Lehtinen et~al.}(2016)\citenamefont{Lehtinen, Naydenov,
  B\"orner, Melentjevic, M\"uller, McGuinness, Pezzagna, Meijer, Kaiser, and
  Jelezko}}]{Leh16}
\bibinfo{author}{\bibfnamefont{O.}~\bibnamefont{Lehtinen}},
  \bibinfo{author}{\bibfnamefont{B.}~\bibnamefont{Naydenov}},
  \bibinfo{author}{\bibfnamefont{P.}~\bibnamefont{B\"orner}},
  \bibinfo{author}{\bibfnamefont{K.}~\bibnamefont{Melentjevic}},
  \bibinfo{author}{\bibfnamefont{C.}~\bibnamefont{M\"uller}},
  \bibinfo{author}{\bibfnamefont{L.~P.} \bibnamefont{McGuinness}},
  \bibinfo{author}{\bibfnamefont{S.}~\bibnamefont{Pezzagna}},
  \bibinfo{author}{\bibfnamefont{J.}~\bibnamefont{Meijer}},
  \bibinfo{author}{\bibfnamefont{U.}~\bibnamefont{Kaiser}}, \bibnamefont{and}
  \bibinfo{author}{\bibfnamefont{F.}~\bibnamefont{Jelezko}},
  \bibinfo{journal}{Phys. Rev. B} \textbf{\bibinfo{volume}{93}},
  \bibinfo{pages}{035202} (\bibinfo{year}{2016}).

\bibitem[{\citenamefont{Tetienne et~al.}(2018)\citenamefont{Tetienne, de~Gille,
  Broadway, Teraji, Lillie, McCoey, Dontschuk, Hall, Stacey, Simpson
  et~al.}}]{Tet18}
\bibinfo{author}{\bibfnamefont{J.~P.} \bibnamefont{Tetienne}},
  \bibinfo{author}{\bibfnamefont{R.~W.} \bibnamefont{de~Gille}},
  \bibinfo{author}{\bibfnamefont{D.~A.} \bibnamefont{Broadway}},
  \bibinfo{author}{\bibfnamefont{T.}~\bibnamefont{Teraji}},
  \bibinfo{author}{\bibfnamefont{S.~E.} \bibnamefont{Lillie}},
  \bibinfo{author}{\bibfnamefont{J.~M.} \bibnamefont{McCoey}},
  \bibinfo{author}{\bibfnamefont{N.}~\bibnamefont{Dontschuk}},
  \bibinfo{author}{\bibfnamefont{L.~T.} \bibnamefont{Hall}},
  \bibinfo{author}{\bibfnamefont{A.}~\bibnamefont{Stacey}},
  \bibinfo{author}{\bibfnamefont{D.~A.} \bibnamefont{Simpson}},
  \bibnamefont{et~al.}, \bibinfo{journal}{Physical Review B}
  \textbf{\bibinfo{volume}{97}}, \bibinfo{pages}{11} (\bibinfo{year}{2018}).

\bibitem[{\citenamefont{Healey et~al.}(2020)\citenamefont{Healey, Stacey,
  Johnson, Broadway, Teraji, Simpson, Tetienne, and Hollenberg}}]{Hea20}
\bibinfo{author}{\bibfnamefont{A.~J.} \bibnamefont{Healey}},
  \bibinfo{author}{\bibfnamefont{A.}~\bibnamefont{Stacey}},
  \bibinfo{author}{\bibfnamefont{B.~C.} \bibnamefont{Johnson}},
  \bibinfo{author}{\bibfnamefont{D.~A.} \bibnamefont{Broadway}},
  \bibinfo{author}{\bibfnamefont{T.}~\bibnamefont{Teraji}},
  \bibinfo{author}{\bibfnamefont{D.~A.} \bibnamefont{Simpson}},
  \bibinfo{author}{\bibfnamefont{J.-P.} \bibnamefont{Tetienne}},
  \bibnamefont{and} \bibinfo{author}{\bibfnamefont{L.~C.~L.}
  \bibnamefont{Hollenberg}}, \bibinfo{journal}{Phys. Rev. Materials}
  \textbf{\bibinfo{volume}{4}}, \bibinfo{pages}{104605} (\bibinfo{year}{2020}).

\bibitem[{\citenamefont{Dobrovitski et~al.}(2009)\citenamefont{Dobrovitski,
  Feiguin, Hanson, and Awschalom}}]{Dob09}
\bibinfo{author}{\bibfnamefont{V.~V.} \bibnamefont{Dobrovitski}},
  \bibinfo{author}{\bibfnamefont{A.~E.} \bibnamefont{Feiguin}},
  \bibinfo{author}{\bibfnamefont{R.}~\bibnamefont{Hanson}}, \bibnamefont{and}
  \bibinfo{author}{\bibfnamefont{D.~D.} \bibnamefont{Awschalom}},
  \bibinfo{journal}{Physical Review Letters} \textbf{\bibinfo{volume}{102}},
  \bibinfo{pages}{237601} (\bibinfo{year}{2009}).

\bibitem[{\citenamefont{Hall et~al.}(2014)\citenamefont{Hall, Cole, and
  Hollenberg}}]{Hal14}
\bibinfo{author}{\bibfnamefont{L.~T.} \bibnamefont{Hall}},
  \bibinfo{author}{\bibfnamefont{J.~H.} \bibnamefont{Cole}}, \bibnamefont{and}
  \bibinfo{author}{\bibfnamefont{L.~C.~L.} \bibnamefont{Hollenberg}},
  \bibinfo{journal}{Phys. Rev. B} \textbf{\bibinfo{volume}{90}},
  \bibinfo{pages}{075201} (\bibinfo{year}{2014}).

\bibitem[{\citenamefont{Lang et~al.}(2019{\natexlab{b}})\citenamefont{Lang,
  Madhavan, Tetienne, Broadway, Hall, Teraji, Monteiro, Stacey, and
  Hollenberg}}]{Lan18}
\bibinfo{author}{\bibfnamefont{J.~E.} \bibnamefont{Lang}},
  \bibinfo{author}{\bibfnamefont{T.}~\bibnamefont{Madhavan}},
  \bibinfo{author}{\bibfnamefont{J.-P.} \bibnamefont{Tetienne}},
  \bibinfo{author}{\bibfnamefont{D.~A.} \bibnamefont{Broadway}},
  \bibinfo{author}{\bibfnamefont{L.~T.} \bibnamefont{Hall}},
  \bibinfo{author}{\bibfnamefont{T.}~\bibnamefont{Teraji}},
  \bibinfo{author}{\bibfnamefont{T.~S.} \bibnamefont{Monteiro}},
  \bibinfo{author}{\bibfnamefont{A.}~\bibnamefont{Stacey}}, \bibnamefont{and}
  \bibinfo{author}{\bibfnamefont{L.~C.~L.} \bibnamefont{Hollenberg}},
  \bibinfo{journal}{Phys. Rev. A} \textbf{\bibinfo{volume}{99}},
  \bibinfo{pages}{012110} (\bibinfo{year}{2019}{\natexlab{b}}).

\bibitem[{\citenamefont{Fávaro~de Oliveira
  et~al.}(2017)\citenamefont{Fávaro~de Oliveira, Antonov, Wang, Neumann,
  Momenzadeh, Häußermann, Pasquarelli, Denisenko, and Wrachtrup}}]{Fav17}
\bibinfo{author}{\bibfnamefont{F.}~\bibnamefont{Fávaro~de Oliveira}},
  \bibinfo{author}{\bibfnamefont{D.}~\bibnamefont{Antonov}},
  \bibinfo{author}{\bibfnamefont{Y.}~\bibnamefont{Wang}},
  \bibinfo{author}{\bibfnamefont{P.}~\bibnamefont{Neumann}},
  \bibinfo{author}{\bibfnamefont{S.~A.} \bibnamefont{Momenzadeh}},
  \bibinfo{author}{\bibfnamefont{T.}~\bibnamefont{Häußermann}},
  \bibinfo{author}{\bibfnamefont{A.}~\bibnamefont{Pasquarelli}},
  \bibinfo{author}{\bibfnamefont{A.}~\bibnamefont{Denisenko}},
  \bibnamefont{and}
  \bibinfo{author}{\bibfnamefont{J.}~\bibnamefont{Wrachtrup}},
  \bibinfo{journal}{Nature Communications} \textbf{\bibinfo{volume}{8}},
  \bibinfo{pages}{15409} (\bibinfo{year}{2017}).

\end{thebibliography}

\end{document}